\documentclass[fleqn,usenatbib]{mnras}

\usepackage{newtxtext,newtxmath}

\usepackage[T1]{fontenc}

\usepackage{graphicx}
\usepackage{amsmath}
\usepackage[export]{adjustbox}

\newcommand{\grb}{GRB~210704A}

\title[\grb{}: A Luminous Fast Blue Transient]{\grb{}: A Luminous Fast Blue Transient in a GRB Afterglow\\ at $z=2.34$}

\author[D.~L.~A. Pieterse et al.]{
\href{https://orcid.org/0000-0003-3114-2733}{Dani\"elle L.~A. Pieterse},$^{1}$\thanks{Email: \href{mailto:d.pieterse@astro.ru.nl}{d.pieterse@astro.ru.nl}}
\href{https://orcid.org/0000-0001-7821-9369}{Andrew J. Levan},$^{1,2}$
\href{https://orcid.org/0000-0003-3193-4714}{Maria E. Ravasio},$^{1,3}$
\href{https://orcid.org/0000-0002-9267-6213}{Jillian C. Rastinejad},$^{4,5}$
\newauthor
\href{https://orcid.org/0009-0005-5404-2745}{Agnes P.~C. van Hoof},$^{1}$
\href{https://orcid.org/0000-0002-7517-326X}{Daniele B. Malesani},$^{6,7,1}$
\href{https://orcid.org/0000-0003-2700-1030}{Nikhil Sarin},$^{8,9}$
\href{https://orcid.org/0000-0001-5169-4143}{Gavin P. Lamb},$^{10}$
\newauthor
\href{https://orcid.org/0000-0001-5108-0627}{Antonio Martin-Carrillo},$^{11}$
\href{https://orcid.org/0000-0002-2028-9329}{Anya E. Nugent},$^{12}$
\href{https://orcid.org/0000-0003-3274-6336}{Nial R. Tanvir},$^{13}$
\href{https://orcid.org/0000-0001-5679-0695}{Peter G. Jonker},$^{1}$
\newauthor
\href{https://orcid.org/0000-0003-2902-3583}{David Alexander Kann},$^{14,15}$\thanks{Deceased}
\href{https://orcid.org/0000-0001-6991-7616}{Jos\'e Feliciano Ag\"u\'i Fern\'andez},$^{16}$
\href{https://orcid.org/0000-0002-9392-9681}{Edo Berger},$^{12}$
\href{https://orcid.org/0009-0009-1573-8300}{Gregory Corcoran},$^{11}$
\newauthor
\href{https://orcid.org/0000-0003-2910-6565}{Felice Cusano},$^{17}$
\href{https://orcid.org/0000-0001-7164-1508}{Paolo D'Avanzo},$^{3}$
\href{https://orcid.org/0000-0002-7320-5862}{Valerio D'Elia},$^{18}$
\href{https://orcid.org/0000-0001-7717-5085}{Antonio de Ugarte Postigo},$^{19}$
\href{https://orcid.org/0000-0001-9868-9042}{Dimple},$^{20,21}$
\newauthor
\href{https://orcid.org/0000-0002-7374-935X}{Wen-fai Fong},$^{4}$
\href{https://orcid.org/0000-0002-8149-8298}{Johan P.U. Fynbo},$^{6,7}$
\href{https://orcid.org/0000-0001-9695-8472}{Luca Izzo},$^{22,23}$
\href{https://orcid.org/0000-0003-2593-4355}{Elisabetta Maiorano},$^{17}$
\href{https://orcid.org/0000-0002-2810-2143}{Andrea Melandri},$^{24}$
\newauthor
\href{https://orcid.org/0000-0002-8691-7666}{Eliana Palazzi},$^{17}$
\href{https://orcid.org/0000-0001-8602-4641}{Jonathan Quirola-V\'asquez},$^{1}$
\href{https://orcid.org/0000-0002-8860-6538}{Andrea Rossi}$^{17}$
and \href{https://orcid.org/0000-0003-3937-0618}{Alicia Rouco Escorial}$^{25}$\\
\textit{Affiliations are listed at the end of the paper}
}

\date{Accepted 18 March 2026.}

\pubyear{\the\year{}}

\begin{document}
\label{firstpage}
\pagerange{\pageref{firstpage}--\pageref{lastpage}}
\maketitle

\begin{abstract}
We present detailed, multi-wavelength analysis of \grb{}: a {\em Fermi} Gamma-ray Burst Monitor discovered and {\em Fermi} Large Area Telescope (LAT) detected gamma-ray burst (GRB). The burst is dominated by a short ($\approx 2$\,s) pulse followed by weaker, softer emission. We line stack our afterglow spectrum and determine the most likely redshift to be $z=2.34$. This is corroborated by the photometric redshift of the extended source underlying the GRB. The spectral energy distribution fit parameters, late-time imaging, as well as the GRB's energetics, spectral lag, and location point to a collapsar nature. Follow-up observations reveal excess optical/infrared emission with respect to a standard afterglow, peaking around $T_0+7$\,d ($2$\,d in the rest frame). The excess is extremely luminous ($M_r=-22.0$\,mag) and rapidly evolving. Strikingly, it resembles the emission seen in recently discovered {\em Einstein Probe} fast X-ray transients EP241021a and EP240414a, as well as the population of luminous fast blue optical transients (LFBOTs). This provides a link between these sources and GRBs. {\em Fermi}/LAT observations imply a high Lorentz factor, making this a case where LFBOT-like emission is also associated with a powerful successfully launched jet. 
We model the excess as likely coming from an energetic refreshed shock.
\end{abstract}



\section{Introduction}

Gamma-ray bursts (GRBs) are the most energetic explosions in the Universe. GRBs are categorised into two classes based on their prompt emission duration~\citep{norris84, kouveliotou93}: long GRBs (LGRBs; $T_{90}>2$\,s) and short GRBs (SGRBs). Here $T_{90}$ is the duration in which $5$--$95$\,per cent of the burst fluence is observed. Although this division is empirical, it is often employed to reflect a physical distinction in progenitor systems. LGRBs are commonly associated with the core-collapse of massive stars. This is evidenced by accompanying broad-lined Type Ic supernovae (SNe Ic-BL; e.g. \citealt{galama98, hjorth03}) and associations to star-forming, low-mass, and low-metallicity galaxies~\citep{fruchter06, perley13, palmerio19}. In contrast, SGRBs are typically attributed to compact object mergers involving neutron stars, where the radioactive decay of neutron-rich ejecta powers a kilonova (e.g. \citealt{tanvir13, abbott17, goldstein17, metzger19}). SGRBs are connected to a variety of host galaxies, including quiescent hosts, and can be found at larger offsets due to natal kicks and in more massive hosts with a broader range of metallicities~\citep{levan16, fong22, nugent22, oconnor22}.

This classification scheme is enriched by the subclass of SGRBs with extended emission (EE-GRBs; \citealt{norris06}). In these bursts, the initial short, hard gamma-ray spike is followed by longer-lasting, softer emission (e.g. \citealt{kaneko15}). EE-GRBs are typically associated to compact object mergers~\citep{gompertz20}, just like the SGRBs without extended emission. They also have host galaxies that are indistinguishable from the population of other SGRB hosts~\citep{fong22}. The physical origin of the extended emission remains debated. Proposed explanations include magnetar spin-down~\citep{metzger08, bucciantini12}, fallback accretion~\citep{rosswog07, metzger10}, multiple jet components~\citep{barkov11}, and magnetic reconnection in the relativistic GRB jet~\citep{zhang11}.

Recent discoveries have challenged the paradigm that LGRBs are created in collapsars and SGRBs in mergers. GRB~200826A~\citep{ahumada21, rossi22} was a short burst that was accompanied by an SN, revealing a collapsar nature. More markedly, LGRBs 211211A~\citep{rastinejad22, troja22, yang22} and 230307A~\citep{levan24, yang24} displayed compelling kilonova signatures, strongly suggesting compact object merger origins. Several more archival events can subsequently be added to this sample, including LGRBs 060614 \citep{gehrels06, jin15, yang15} and 191019A~\citep{levan23, stratta25}. These cases demonstrate that both LGRBs and SGRBs can emerge from diverse progenitors and that there is overlap between their formation channels.

\grb{} was a GRB with a short initial spike, followed by softer, extended emission up to about $6$\,s~\citep{becerra23}. It generated considerable interest within the astronomical community due to its borderline duration and possible EE-GRB classification, its high-energy GeV gamma-ray detections~\citep{berretta21_gcn}, and its location in an apparent over-density of luminous galaxies. The latter is of interest as these galaxies form plausible host galaxies for compact object merger progenitors~\citep{levan21_gcn}. Follow-up observations revealed luminous optical/infrared (IR) excess emission with respect to a standard afterglow $5$\,d after the burst, prompting further questions about \grb{}'s physical origin.

Typically, the interaction of the GRB jet with the circumstellar medium (CSM) results in afterglow emission that decays as a power law with a slope of $\alpha\approx-1.2$~\citep{zhang06}. Rebrightening phases have been observed hours to days after GRB explosions. They can be caused by e.g. the emergence of thermal transients such as SNe or kilonovae, non-uniform jet structures viewed off-axis, density variations in the CSM~\citep{wang00, dai02, lazzati02, nakar03}, SN-CSM interaction~(e.g. \citealt{fraser20, margalit22}), or prolonged central-engine activity. The latter includes refreshed shocks, where the central engine ejects shells of relativistic material which can cause a rebrightening when a slower, more energetic shell of ejecta catches up and re-energises the decelerating shock front~\citep{rees98, sari00, zhang02, lamb19, anderson25}.

\citet{becerra23} highlight that the brightness and timing of \grb{}'s excess emission are difficult to explain with standard kilonova and SN scenarios. They suggest an exotic merger channel involving a white dwarf and a compact object (white dwarf, neutron star or black hole) in a galaxy cluster environment at $z=0.2$. However, they note that this is a poorly constrained scenario and no detailed modelling was performed by them.

In this work, we present multi-wavelength observations of \grb{}, including those of our own observing campaign. We extend the data set presented by \citet{becerra23} with additional $R/r$- and $K$-band detections and with the {\em Fermi} Large Area Telescope (LAT) data (Sec.~\ref{sec:obs}). We perform independent photometry and data analysis. We line stack our afterglow spectrum and perform spectral energy distribution (SED) fitting, which indicate a redshift of $z=2.34$. Our analysis is presented in Sec.~\ref{sec:results}. Additionally, we compare \grb{} to other transients, and we use Bayesian inference package {\sc redback} to model the afterglow and the observed excess emission (Sec.~\ref{sec:discussion}). Finally, we summarise our findings and implications in Sec.~\ref{sec:conclusion}.

All magnitudes in this work are in the AB~system and upper limits are $3\sigma$ unless specified differently. We use a flat $\Lambda$CDM cosmology with $H_0 = 66.7$\,km\,s$^{-1}$\,Mpc$^{-1}$ and $\Omega_{\text{M}} = 0.31$~\citep{planck18}. Observation properties are given in the observer frame unless otherwise specified.

\section{Observations}\label{sec:obs}

\subsection{Gamma Rays}

\begin{figure}
    \centering
    \includegraphics[width=0.47\textwidth]{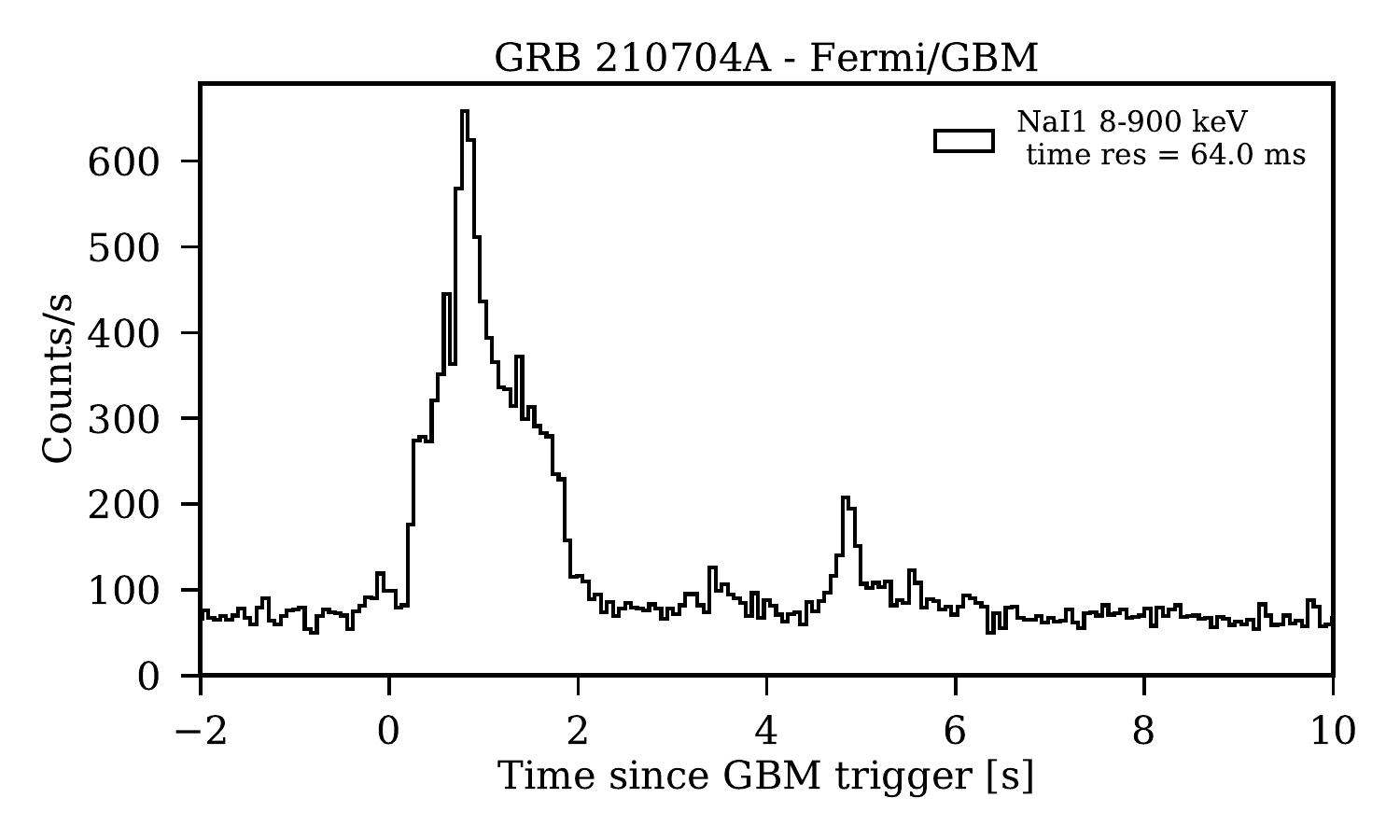}\\
    \includegraphics[width=0.44\textwidth]{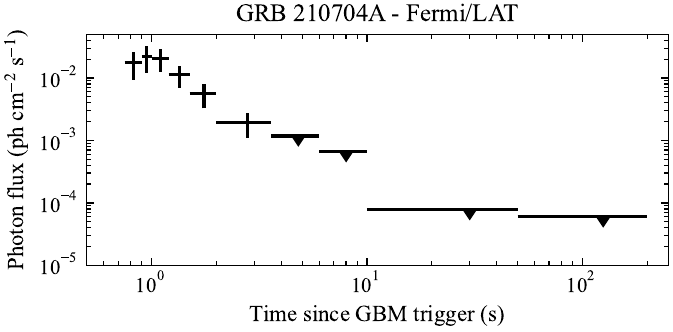}
    \caption{Top panel: {\em Fermi}/GBM light curve of \grb{}, spanning $8$--$900$\,keV and using $64$\,ms bins (not background-subtracted). Bottom panel: {\em Fermi}/LAT light curve of \grb{}.}
    \label{fig:gbm_lc}
\end{figure}

\grb{} was discovered by the {\em Fermi} Gamma-ray Burst Monitor (GBM; \citealt{meegan09}) on 2021 July 4 at 19:33:24.59 UT (trigger time $T_0$), with a fluence of $(1.95 \pm 0.02) \times 10^{-5}$\,erg\,cm$^{-2}$ ($10$--$1000$\,keV and $0.2$--$1.8$\,s post-trigger; \citealt{malacaria21_gcn}). Its prompt emission was also detected by {\em AGILE}~\citep{ursi21_gcn}, {\em AstroSat}~\citep{prasad21_gcn}, {\em INTEGRAL}~\citep{minaev21_gcn}, and {\em Konus-Wind}~\citep{ridnaia21_gcn}. The {\em Fermi}/GBM light curve in Fig.~\ref{fig:gbm_lc} exhibits a primary emission peak lasting for $2$\,s, followed by fainter extended emission peaking around $4.5$\,s. \citet{ridnaia21_gcn} also observe this light curve behaviour with {\em Konus-Wind}. The extended emission results in a $T_{90}$ of $4.7$\,s ($50$--$300$\,keV), $4.6^{+2.8}_{-3.6}$\,s ($20$--$200$\,keV), and $3.5\pm0.7$\,s ($80$--$8000$\,keV) for {\em Fermi}/GBM~\citep{malacaria21_gcn}, {\em AstroSat}~\citep{becerra23}, and {\em INTEGRAL}~\citep{minaev21_gcn}, respectively. This formally places the burst in the long-duration GRB population \citep{kouveliotou93}, although more accurately it can be classified as an EE-GRB. \citet{becerra23} demonstrate that the extended emission is soft ($<300$\,keV) compared to the main emission peak. This explains the short duration ($1.06$\,s) reported by \citet{ursi21_gcn} for the {\em AGILE} observation in its higher-energy band of $400$\,keV--$100$\,MeV.

The spectral lag between the energy bands $25$--$50$\,keV and $100$--$300$\,keV is $\tau=80\pm9$\,ms~\citep{becerra23}.

We analysed the GBM data from the two NaI detectors and the BGO detector with the best viewing angle to the source (n1, n5, and b0). Spectral data files and the corresponding most updated response matrix files (rsp2) are obtained from the online archive.\footnote{\url{https://heasarc.gsfc.nasa.gov/W3Browse/fermi/fermigbrst.html}} For the time-integrated analysis of the burst ($0$--$4$\,s from GBM trigger time), we made use of CSPEC data, which have $1024$\,ms time resolution. We selected the energy channels in the range $10$--$900$\,keV for NaI detectors and $0.3$--$40$\,MeV for BGO detectors, and we exclude the channels in the range $30$--$40$\,keV due to the presence of the Iodine K-edge at $33.17$\,keV.\footnote{\url{https://fermi.gsfc.nasa.gov/ssc/data/analysis/GBM\_caveats.html}} Spectra were extracted with the public software {\sc gtburst}.\footnote{\url{https://fermi.gsfc.nasa.gov/ssc/data/analysis/scitools/gtburst.html}} We discuss the prompt emission spectrum in Sec.~\ref{sec:fermiobs}.

High-energy gamma rays of \grb{} were detected by the {\em Fermi}/LAT~\citep{atwood09}. The LAT is a pair-conversion telescope covering the energy range from $30$\,MeV to more than $300$\,GeV. For the extraction and analysis of LAT data, we use {\sc gtburst}, following the procedure described in the online official {\em Fermi} guide\footnote{\url{https://fermi.gsfc.nasa.gov/ssc/data/analysis/scitools/gtburst.html}} and performing an unbinned likelihood analysis.

We select P8R3\_TRANSIENT020 class events, filtering photons in the $100$\,MeV -- $20$\,GeV energy range from a region of interest (ROI) of $12^\circ$ radius centred on the burst location. We apply a maximum zenith angle cut of $100^\circ$ to reduce contamination of gamma-rays from the Earth limb.

We analysed {\em Fermi}/LAT data over the first 200\,s from the GBM trigger time. A time-resolved analysis was performed to characterise the emission at GeV energies and track its temporal evolution. The source is significantly detected (test statistic (TS) $>10$, corresponding to $\approx 3 \sigma$) only in the interval $0.75$--$3.6$\,s, with TS between $60$--$119$. At earlier times ($0$--$0.75$\,s) and after $t > T_0+3.6$\,s, the source is not detected, hence we derived upper limits on the flux. When the source is not detected, the upper limits were computed at 95\% confidence level and assuming a photon index fixed to $-2$. The {\em Fermi}/LAT light curve of \grb{} is presented in Fig.~\ref{fig:gbm_lc}. The reported photon flux above $100$\,MeV up to $T_0+10$\,s is $F_{\text{ph}}=(1.60 \pm 0.31) \times 10^{-3}$\,ph\,cm$^{-2}$\,s$^{-1}$~\citep{berretta21_gcn}.

\subsection{Follow-up}

{\em Fermi}/LAT and the {\em Swift} X-Ray Telescope (XRT; \citealt{burrows05}) facilitated localisation efforts~\citep{berretta21_gcn, dai21_gcn}, but it ultimately took $0.94$\,d for the afterglow position to be determined to sub-arcsec precision. This was managed by the $1.5$-m AZT-20 telescope of the Assy-Turgen observatory, which determined the location of \grb{} to be R.A., decl.\,$=\,10^{\text{h}} 36^{\text{m}} 04\fs 88, +57^\circ 12' 58\farcs 5$ in the J2000 frame with an uncertainty of $0.5$\,arcsec~\citep{kim21_gcn}. Subsequently, we initiated a series of imaging observations with a diverse range of ground-based telescopes, encompassing both visible and IR wavelengths, in addition to triggering afterglow spectroscopy. Furthermore, we initiated target of opportunity (ToO) imaging with the {\em Hubble Space Telescope (HST)}. Some of the data have already been reported by \citet{becerra23}, but we perform our own independent analysis. All follow-up efforts are detailed below.

\subsubsection{X-Rays}

{\em Swift}/XRT observed \grb{} at four epochs, between $0.6$--$4.5$\,d after the GBM trigger. We downloaded the high-level data products from the UK Swift Science Data Centre (UKSSDC), which were processed via the techniques described in \citet{evans07, evans09}. From a time-averaged XRT spectrum spanning $T_0+53$\,ks to $T_0+72$\,ks, the UKSSDC reports a photon index of $\Gamma = 1.5^{+0.6}_{-0.4}$. Using this $\Gamma$ and a Galactic column density of $N_{\text{H}} = 5.59 \times 10^{19}$\,cm$^{-2}$, we correct the fluxes for Galactic extinction using Chandra's Portable, Interactive Multi-Mission Simulator~\citep{mukai93}.

An additional X-ray observation epoch ($T_0+14.4$\,d) was obtained with the {\em Chandra X-ray Observatory} (proposal 22508794, PI: Troja) on 2021 July 19 with the ACIS-S array. The afterglow is detected with $12$~counts and a count rate of $(6.6^{+2.1}_{-1.7}) \times 10^{-4}$ counts~s$^{-1}$, translating to an unabsorbed flux of $F_{\text{X}} = (1.3^{+0.4}_{-0.3}) \times 10^{-14}$~erg~cm$^{-2}$\,s$^{-1}$ ($0.3$--$10$\,keV; \citealt{becerra23}).

\subsubsection{Visible Bands}

For our ground-based follow-up campaign, we initiated visible-band imaging utilizing the Device Optimized for the LOw RESolution (DOLORES) on the $3.6$-m Telescopio Nazionale Galileo (TNG; \citealt{davanzo21_gcn_1, davanzo21_gcn_2}), the $0.8$-m T80 at the Observatorio Astrof\'isico de Javalambre (OAJ; \citealt{kann21_gcn_oaj}), the Calar Alto Faint Object Spectrograph (CAFOS) on the $2.2$-m telescope at the Calar Alto Observatory (CAHA; \citealt{kann21_gcn_caha}), the Optical System for Imaging and low-intermediate-Resolution Integrated Spectroscopy (OSIRIS) mounted on the $10.4$-m Gran Telescopio Canarias (GTC; \citealt{deugartepostigo21_gcn}) and the Alhambra Faint Object Spectrograph and Camera (AlFOSC) on the $2.6$-m Nordic Optical Telescope (NOT; \citealt{kann21_gcn_not}). We include additional images taken with GTC/OSIRIS~\citep{watson21_gcn}, NOT/ALFOSC, TNG/DOLORES, the Gemini Multi-Object Spectrographs (GMOS) on the $8.1$-m Gemini North telescope and the Tiny Wide Field Camera~2 (TWFC2) on the $4.2$-m William Herschel Telescope (WHT).

To obtain homogeneous photometry, we process all aforementioned optical observations with our custom photometry pipeline. We use {\sc astrometry.net}~\citep{lang10} for the astrometric calibration and {\sc sextractor}~\citep{bertin96} for background estimation and source extraction. Using the {\sc photutils} package~\citep{bradley24}, we perform aperture photometry on the background-subtracted images. We use Pan-STARRS DR2~\citep{flewelling20} for the photometric calibration, selecting six calibration sources when available. We note that we could only use one or two calibration sources for the data from Gemini/GMOS and GTC. All photometry is presented in Table~\ref{tab:photometry_optIR}.

We supplement our data with photometry from General Coordinates Network reports, taken with the {\em Swift} Ultra-Violet/Optical Telescope (UVOT; \citealt{roming05}), the AZT-20 telescope of the Assy-Turgen observatory, the Reionization And Transients IR camera (RATIR) on the Harold Johnson Telescope (HJT) at the Observatorio Astron\'omico Nacional San Pedro M\'artir (OAN-SPM), the $2.2$-m telescope at CAHA, the Large Monolithic Imager (LMI) on the Lowell Discovery Telescope (LDT), the Zeiss-1000 telescope of the Special Astrophysical Observatory of the Russian Academy of Science (SAO RAS), GMOS on Gemini North and the Hyper Suprime-Cam (HSC) on the Subaru telescope. \citet{dichiara21_gcn} also obtained a $5\sigma$-upper limit with the Deca-Degree Optical Transient Imager (DDOTI) at the OAN-SPM, which we convert to a $3\sigma$-limiting magnitude with an additional term of $-2.5\, \text{log}_{10}(3/5)$. This supplemental photometry is listed in Table~\ref{tab:photometry_optIR} along with the appropriate references.

\begin{table*}
    \centering
    \caption{Post-GRB photometry, not corrected for extinction. Listed are the time since the burst (using the mid-exposure time) in days, the exposure time in seconds, the telescope/instrument, the observation filter, the AB magnitude of the detection or $3\sigma$ upper limit (indicated with the $>$-symbol) and the reference for the photometry.}
    \label{tab:photometry_optIR}
    \begin{tabular}{ccllcl}
        \hline\hline
        $t_{\text{mid}}-T_0$ (d) &
        $t_{\text{exp}}$ (s) &
        Instrument &
        Band & 
        Magnitude (AB) &
        Reference \\
        \hline
        $0.391$   & $360$   & DDOTI             & $w$   & $>20.3$ & \citet{dichiara21_gcn}\\
        $0.726$   & $3038$  & {\em Swift}/UVOT  & $V$   & $>21.1$ & \citet{breeveld21_gcn}\\
        $0.944$   & $3960$  & AZT-20            & $r'$  & $22.25 \pm 0.13$ & \citet{kim21_gcn}\\
        $1.080$   & $1800$  & TNG/DOLORES       & $r$   & $22.50 \pm 0.07$ & This work\\
        $1.081$   & $900$   & OAJ/T80Cam        & $z'$  & $>21.4$ & This work\\
        $1.093$   & $1500$  & CAHA-2.2m/CAFOS   & $i$   & $21.66 \pm 0.29$ & This work\\
        $1.093$   & $600$   & OAJ/T80Cam        & $i'$  & $22.26 \pm 0.46$ & This work\\
        $1.094$   & $60$    & GTC/OSIRIS        & $r'$  & $22.24 \pm 0.09$ & This work\\
        $1.099$   & $1200$  & WHT/TWFC2         & $r$   & $22.74 \pm 0.05$ & This work\\
        $1.104$   & $900$   & OAJ/T80Cam        & $g'$  & $22.84 \pm 0.29$ & This work\\ 
        $1.113$   & $1200$  & WHT/TWFC2         & $z$   & $>22.6$ & This work\\
        $1.116$   & $900$   & OAJ/T80Cam        & $r'$  & $22.63 \pm 0.26$ & This work\\
        $1.29$    & $1152$  & HJT/RATIR         & $r$   & $>22.4$ & \citet{becerra23}\\
        $1.29$    & $1152$  & HJT/RATIR         & $i$   & $>22.1$ & \citet{becerra23}\\
        $2.062$   & $2700$  & CAHA-2.2m/CAFOS   & $r$   & $>22.7$ & \citet{sun21_gcn}\\ 
        $2.096$   & $1200$  & WHT/TWFC2         & $z$   & $>22.6$ & This work\\
        $2.104$   & $420$   & GTC/OSIRIS        & $r'$  & $23.63 \pm 0.11$ & This work\\
        $2.107$   & $540$   & GTC/OSIRIS        & $z'$  & $23.69 \pm 0.46$ & This work\\
        $2.114$   & $1600$  & WHT/TWFC2         & $r$   & $23.33 \pm 0.11$ & This work\\
        $3.920$   & $4620$  & AZT-20            & $r'$   & $>23.1$ & \citet{pankov21_gcn}\\
        $4.095$   & $2700$  & TNG/DOLORES       & $i$   & $24.08 \pm 0.28$ & This work\\
        $4.122$   & $900$   & TNG/DOLORES       & $r$   & $23.83 \pm 0.30$ & This work\\
        $4.457$   & $840$   & Gemini North/NIRI & $K$   & $23.18 \pm 0.08$ & This work\\
        $5.117$   & $1080$  & GTC/OSIRIS        & $r'$  & $23.48 \pm 0.08$ & This work\\
        $5.119$   & $840$   & GTC/OSIRIS        & $z'$  & $22.59 \pm 0.17$ & This work\\
        $5.447$   & $1440$  & Gemini North/NIRI & $J$   & $22.88 \pm 0.21$ & This work\\
        $5.474$   & $1980$  & Gemini North/NIRI & $K$   & $23.35 \pm 0.23$ & This work\\
        $6.086$   & $900$   & NOT/ALFOSC        & $g'$  & $23.61 \pm 0.17$ & This work\\
        $6.097$   & $900$   & NOT/ALFOSC        & $r'$  & $23.52 \pm 0.15$ & This work\\
        $6.35$    & $1500$  & LDT/LMI           & $r$   & $23.14 \pm 0.27$ & \citet{becerra23}\\
        $6.451$   & $2220$  & Gemini North/NIRI & $K$   & $23.07 \pm 0.10$ & This work\\
        $6.983$   & $3600$  & Zeiss-1000        & $R$   & $23.1 \pm 0.3$ & \citet{volnova21_gcn}\\
        $7.468$   & $660$   & Gemini North/NIRI & $K$   & $22.53 \pm 0.29$ & This work\\
        $8.946$   & $7200$  & Zeiss-1000        & $R$   & $>23.5$ & \citet{volnova21_gcn}\\
        $10.448$  & $1080$  & Gemini North/GMOS & $r$   & $23.16 \pm 0.43$ & This work\\
        $11.44$   & $900$   & Gemini North/GMOS & $z$   & $>22.44$ & \citet{becerra23}\\
        $12.451$  & $1740$  & Gemini North/NIRI & $K$   & $22.73 \pm 0.07$ & This work\\
        $13.091$  & $2100$  & TNG/DOLORES       & $r$   & $>23.3$ & This work\\
        $14.449$  & $1584$  & Gemini North/GMOS & $z$   & $24.10 \pm 0.19$ & This work\\
        $30.081$  & $600$   & NOT/ALFOSC        & $R$   & $>24.0$ & This work\\
        $32.088$  & $900$   & NOT/ALFOSC        & $R$   & $24.18 \pm 0.38$ & This work\\
        $115.00$  & $3030$  & Subaru/HSC        & $r$   & $>25.21$ & \citet{becerra23}\\
        $160.796$ & $7140$  & Gemini North/NIRI & $K$   & $24.65 \pm 0.18$ & This work\\
    \end{tabular}
\end{table*}

\begin{table*}
    \centering
    \caption{Late-time {\em HST} photometry at the location of \grb{}, not corrected for extinction. Listed are the telescope channel, observation band, observation date, time since the GRB (using the start of the exposure) in days, the AB magnitudes using a $1$-arcsec aperture and the AB magnitudes using a $0.2$-arcsec aperture. Upper limits are $2\sigma$ and indicated with the $>$-symbol.}
    \label{tab:photometry_hst}
    \begin{tabular}{llllll}
        \hline\hline
        Channel &
        Band &
        Observation Date &
        $t_{\text{start}}-T_0$ (d) & 
        Magnitude ($1$\,arcsec) &
        Magnitude ($0.2$\,arcsec) \\
        \hline
        UVIS & F336W & 2022 Jan 14 & $193.95$ & $>25.7$ & $26.81 \pm 0.35$ \\
        UVIS & F438W & 2022 Jan 14 & $194.02$ & $>25.4$ & $26.74 \pm 0.25$ \\
        UVIS & F606W & 2022 Jan 14 & $194.08$ & $25.16 \pm 0.17$ & $25.99 \pm 0.04$ \\
        IR   & F105W & 2021 Oct 19 & $107.18$ & $25.06 \pm 0.28$ & $25.69 \pm 0.08$ \\
        IR   & F160W & 2021 Oct 19 & $107.17$ & $24.06 \pm 0.14$ & $24.57 \pm 0.06$ \\
        \hline
    \end{tabular}
\end{table*}

\subsubsection{Infrared}

\grb{} was observed by the Gemini Near-IR Imager (NIRI) mounted on Gemini North, for six epochs in $K$ and one epoch in $J$. We reduce the images using the {\sc dragons} pipeline (Data Reduction for Astronomy from Gemini Observatory North and South; \citealt{labrie23}), employing {\sc iraf} (Image Reduction and Analysis Facility; \citealt{tody86, tody93, fitzpatrick25}) for the astrometry. We then use the {\sc hotpants} routine (High Order Transform of Point-spread function ANd Template Subtraction; \citealt{becker15}) to subtract the last $K$-band epoch ($T_0+161$\,d) from the other $K$-band stacks. For the $K$-band epochs $4$\,d, $6$\,d, and $12$\,d after the burst, this results in a clean image that we use for the photometry instead of the image stack itself. We perform aperture photometry, calibrated to 2MASS (Two Micron All Sky Survey; \citealt{skrutskie06}). The IR photometry is listed in Table~\ref{tab:photometry_optIR}.

\subsubsection{Afterglow Spectrum}

We took $4 \times 600$\,s spectra with OSIRIS mounted on the $10.4$-m GTC telescope at Roque de los Muchachos Observatory, on the island of La Palma (Spain). The spectra were obtained using the R1000B grism, which covers the wavelength range $3700$--$7800$\,\AA{}, as reported by \citet{deugartepostigo21_gcn}. The slit was placed in the parallactic angle to avoid differential slit losses, with a width of $1$\,arcsec, resulting in a spectral resolving power of around $600$. The combined spectrum was obtained at an average epoch of $26.75$\,h after the burst. The spectrum was reduced using self-developed pipelines that include bias and response correction, wavelength calibration, and flux calibration using the spectrophotometric standard Ross~640 as reference. The spectrum is discussed in Sec.~\ref{sec:redshift}.

\subsubsection{Late-time Host Spectrum}

On 2022 February 5 ($216$\,d after the GRB), a set of six spectroscopic exposures of $1200$\,s each were performed with the twin Multi-Object Double Spectrograph MODS-1 and MODS-2~\citep{pogge10} mounted on the Large Binocular Telescope (LBT) on Mt. Graham (Arizona, USA). We used the dual-grating mode, which provided a wavelength coverage of $3200$--$9500$\,\AA{}, and a slit mask with a width of $1.2$\,arcsec. The slit was positioned on the GRB location and oriented with a position angle of $40\fdg 9$ to keep the bright star used for the acquisition in the slit, useful for a precise control of the target placement.

Observations were obtained at an average airmass of $1.5$, seeing $\approx0.8$\,arcsec, and clear sky conditions.

We processed the LBT spectrum using the Spectroscopic Interactive Pipeline and Graphical Interface tool~\citep{gargiulo22} and performed wavelength calibration using arc lamp frames. To obtain flux-calibrated LBT spectra, we applied the sensitivity function derived from the spectrophotometric standard star GD\,71. Only the first $4\times1200$ s exposures were stacked in the blue side ($3200$--$5600$\,\AA{}) because of increasing sky background during morning twilight. Unfortunately, since MODS is not equipped with an atmospheric dispersion corrector (ADC), the signal blueward of $4400$\,\AA{} was lost due to atmospheric dispersion.

\subsubsection{Late-time HST Observations}

Following our ToO ({\em HST} proposal 16275, PI: Tanvir), the field of \grb{} was observed by {\em HST} on 2021 October 19 using the IR channel of Wide Field Camera~3 (WFC3) and on 2022 January 14 with the UVIS channel of WFC3. A log of the observations is shown in Table~\ref{tab:photometry_hst}. The images are reduced via {\sc astrodrizzle}~\citep{avila15} with the final pixel scale set to $0.07$\,arcsec\,pixel$^{-1}$ for the IR channel and $0.025$\,arcsec\,pixel$^{-1}$ for the UVIS channel. For the astrometry, we use our Gemini/NIRI images.

\subsection{Archival Data}

\begin{table}
    \centering
    \caption{Pre-GRB photometry of the source underlying \grb{}, not corrected for extinction. Listed are the telescope/instrument, observation band and AB magnitude. The upper limit is $3\sigma$ and indicated with an $>$-symbol.}
    \label{tab:photometry_archival}
    \begin{tabular}{cllcl}
        \hline\hline
        Telescope &
        Band & 
        Magnitude (AB) \\
        \hline
        CFHT/MegaCam & \textit{u} (MP9301) & $26.51 \pm 0.57$ \\
        CFHT/MegaCam & \textit{g} (MP9401) & $24.99 \pm 0.24$ \\
        CFHT/MegaCam & \textit{r} (MP9601) & $25.38 \pm 0.19$ \\
        CFHT/MegaCam & \textit{i} (MP9701) & $25.48 \pm 0.35$ \\
        CFHT/MegaCam & \textit{z} (MP9801) & $>23.6$ \\
        Subaru/HSC   & \textit{g} & $25.57 \pm 0.08$ \\
        Subaru/HSC   & \textit{i} & $25.54 \pm 0.18$ \\
        Subaru/HSC   & \textit{z} & $25.63 \pm 0.42$ \\
        \hline
    \end{tabular}
\end{table}

The location of \grb{} was observed between 2003 and 2012 -- well before the GRB -- by the Canada-France-Hawaii Telescope (CFHT) in $u$, $g$, $r$, $i$, and $z$ using its MegaPrime imager. Additionally, there are pre-burst archival data taken with the Subaru telescope (in $g$, $i$, and $z$) in the Hyper Suprime-Cam Legacy Archive 2016 (HSCLA; \citealt{tanaka21}). We run the image stacks through our photometry pipeline, where we use SDSS DR19~\citep{sdss25} for the photometric calibration of the $u$-band image and Pan-STARRS DR2~\citep{flewelling20} for the other data. A source is detected at the location of the GRB (Table~\ref{tab:photometry_archival}).

\section{Results}\label{sec:results}


\subsection{{\em Fermi} Observations}\label{sec:fermiobs}

\begin{figure}
    \centering
    \includegraphics[width=0.49\textwidth]{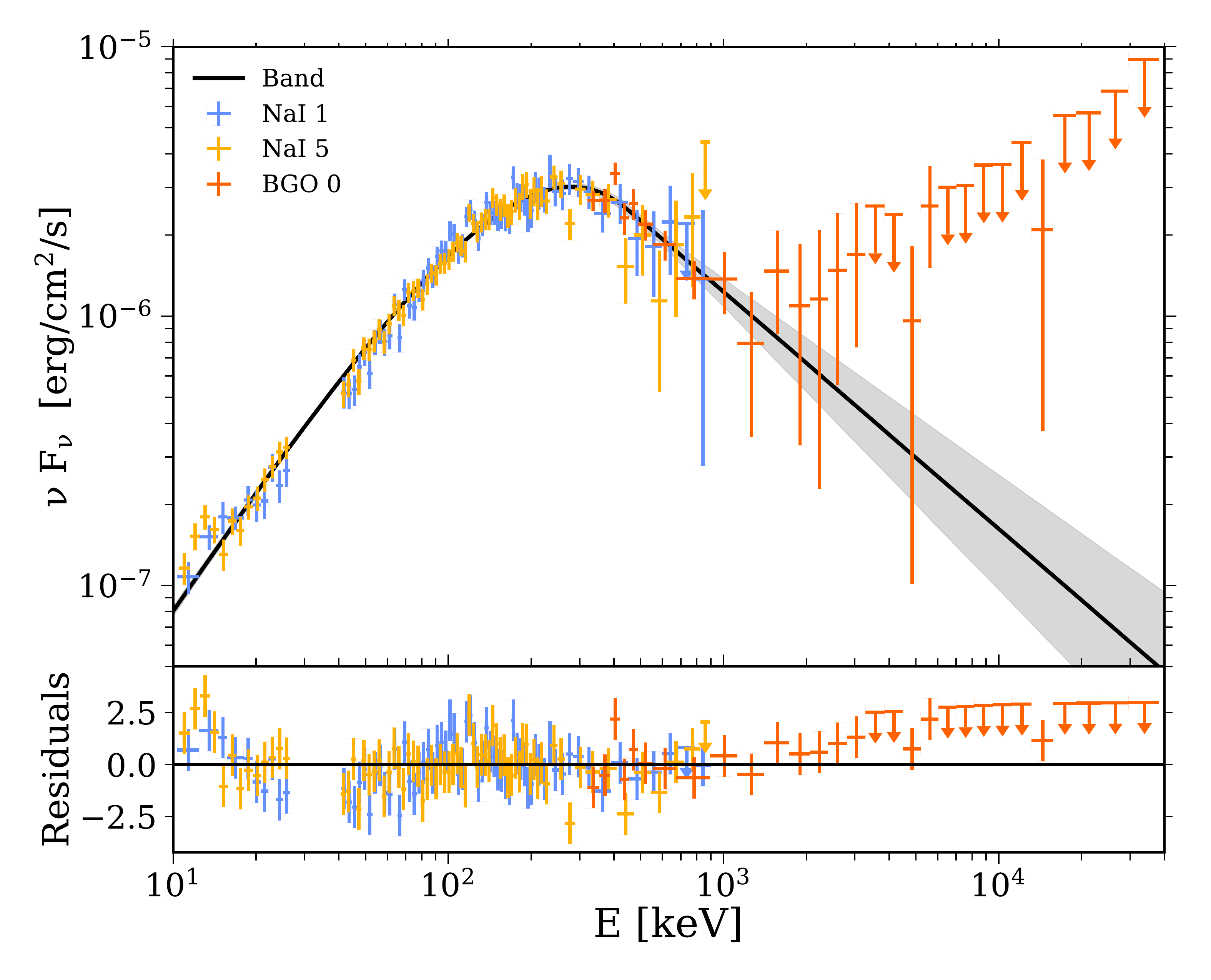}
    \caption{The upper panel shows the {\em Fermi}/GBM spectrum (data points) integrated over the first $4$\,s of \grb{}, overlaid with the best-fitting model (Band; solid graph) and the 16th to 84th percentile model uncertainty interval (shaded region). The lower panel shows the residuals of the fit.} 
    \label{fig:gbm_spectrum}
\end{figure}

We perform a spectral analysis of the prompt {\em Fermi}/GBM data with the public software {\sc xspec}~(v.~12.12.1). We fit a Band function to the {\em Fermi}/GBM energy spectrum of the initial $4$\,s of prompt emission (Fig.~\ref{fig:gbm_spectrum}). We use the PG-Statistic in the fitting procedure, which is valid for Poisson data with a Gaussian background. The fit yields $E_{\text{peak}}=280 \pm 10$\,keV and power-law indices $\alpha=-0.46 \pm 0.03$ and $\beta=-2.86 \pm 0.16$. Analysing the {\em Fermi}/LAT data from the same initial $4$\,s gives a hard power-law spectral index of $-1.64 \pm 0.14$. This hardness contrasts with the significantly softer $\beta$ from the GBM data. Therefore, the LAT emission does not lie on the extrapolation of the lower-energy GBM spectrum, indicating the presence of an additional, harder spectral component at GeV energies. To investigate whether this could be high-energy emission from the GRB afterglow, we include the LAT data in our afterglow modelling in Section~\ref{sec:redback}. Although high-energy afterglow emission is not common, it has been observed before in LAT data (e.g. \citealt{ghirlanda10}).

\begin{figure}
    \centering
    \includegraphics[width=0.4\textwidth]{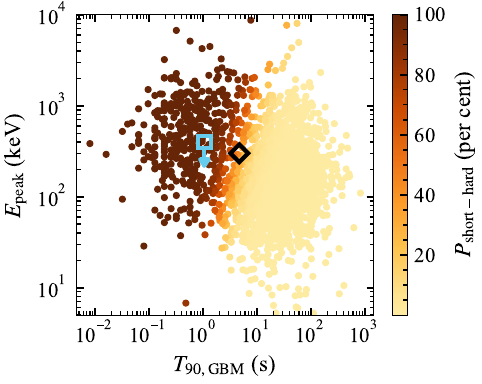}
    \caption{Peak energy vs. duration ($50$--$300$\,keV) for the $2308$ GRBs from the Fourth {\em Fermi}/GBM GRB Catalog~\protect\citep{gruber14, vonkienlin14, narayanabhat16, vonkienlin20}. $E_{\text{peak}}$ is determined with a Band-function fit to a single spectrum over the duration of each burst. The bursts cluster into two groups: short-hard GRBs (top left) and long-soft GRBs (bottom right). Following \protect\citet{ahumada21}, we fit two log-normal distributions to the data to determine the probability of each burst belonging to the cluster of short-hard bursts. This sets the colour scale. \grb{} is marked by a diamond. The square is \grb{} as observed by {\em AGILE}, or in other words, disregarding the extended emission. As the energy peak is not observed in the {\em AGILE} energy range, the square represents an upper limit. All values are in the observer frame.}
    \label{fig:pop_plot_peak}
\end{figure}

We compare the peak energy and duration of \grb{} to the GRB population in Fig.~\ref{fig:pop_plot_peak}. The GRBs in the figure cluster into two groups: short-hard (top left) and long-soft bursts (bottom right). \grb{} (diamond) lies on the transition area between the groups, where it is formally classified as a long-soft burst, as $P_{\text{long-soft}} = 66$\,per cent. Alternatively, using the duration determined with {\em AGILE} ($T_{90}=1.06$\,s in $400$\,keV--$100$\,MeV), which excludes the softer extended emission, \grb{} would be classified as a short-hard burst (square).

\begin{figure}
    \centering
    \includegraphics[width=0.35\textwidth]{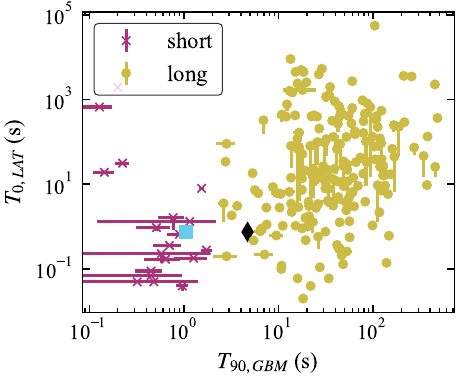}
    \caption{Onset time of the LAT emission ($100$\,MeV--$100$\,GeV) vs. duration ($50$--$300$\,keV) for SGRBs and LGRBs from the Second {\em Fermi}/LAT GRB Catalog~\protect\citep{ajello19}. The onset time is defined as the time when the first photon with probability $>0.9$ of being associated with the GRB is detected in the $100$\,MeV--$100$\,GeV range. \grb{} is marked by a diamond. The square is \grb{} when disregarding the extended emission. All values are in the observer frame.}
    \label{fig:pop_plot_lat}
\end{figure}

The {\em Fermi}/LAT emission of \grb{} is delayed by $0.75$\,s with respect to the {\em Fermi}/GBM emission. Comparing \grb{}'s onset time and duration to other LAT bursts shows that it lies on the periphery of the LGRB population (Fig.~\ref{fig:pop_plot_lat}, diamond). \grb{}'s onset time is representative of the bottom $20$\,per cent of the LAT population, yet is within the range of both SGRBs and LGRBs. When excluding the extended emission, \grb{} is consistent with the SGRB population (square). 

In both figures, the extended emission strongly affects the classification of \grb{}.

\subsection{Environment}\label{sec:environment}

\begin{figure*}
    \centering
    \includegraphics[width=0.7\textwidth]{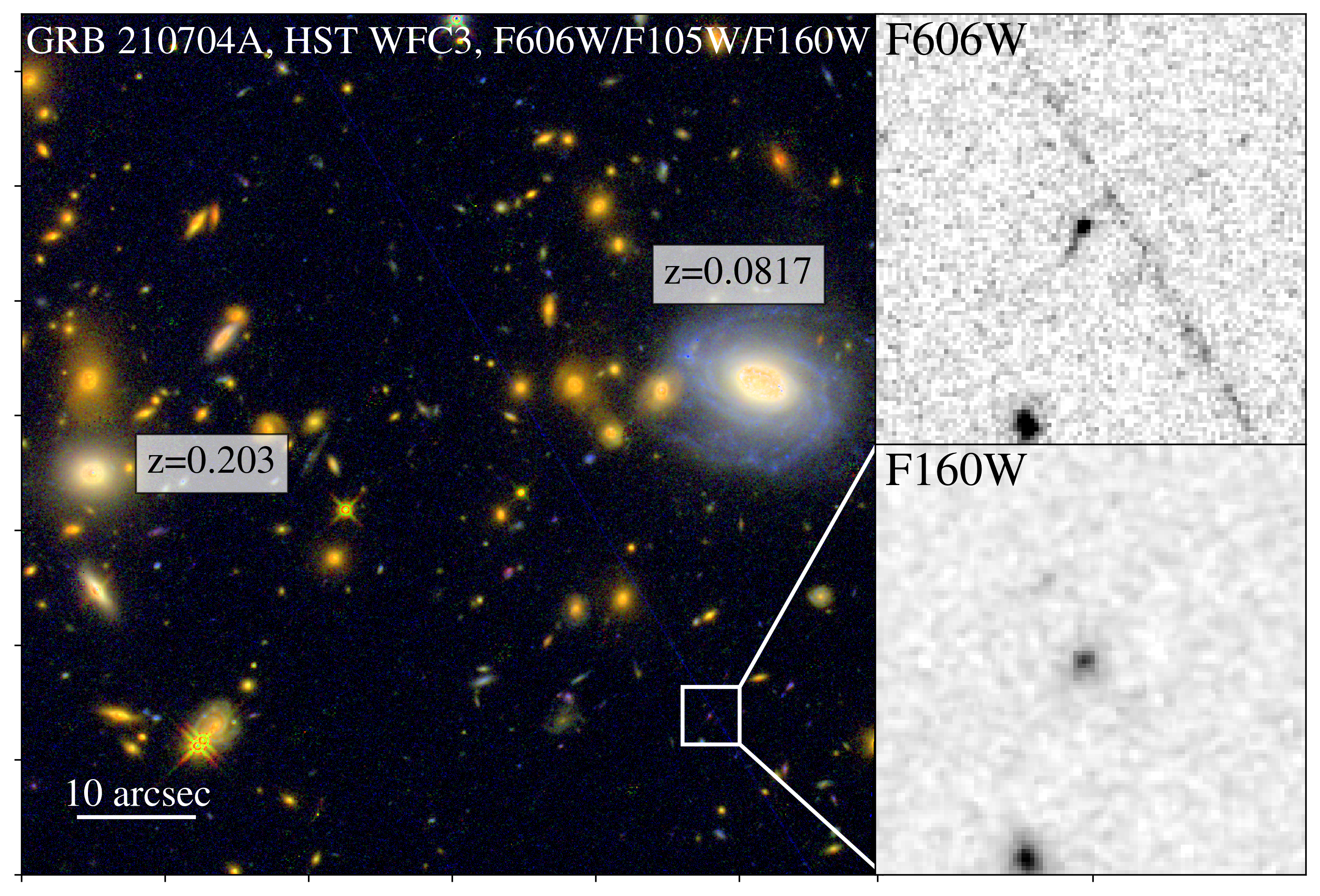}
    \caption{{\bf Left:} Colour-composite image of the environment of \grb{}, made from {\em HST} data. The GRB location is marked by the box. The spiral galaxy on the right side of the image is foreground galaxy WISEA J103604.24$+$571327.7 at a redshift of $z=0.0817$, at a distance of $30$\,arcsec ($45$\,kpc) from \grb{}. The {\em HST} image also overlaps with galaxy cluster 400d J1036$+$5713 at $z=0.203$. {\bf Right:} Zoomed in region around the GRB position showing the galaxy underlying the GRB in the F606W and F160W filters. While in F160W the source is dominated by a compact core, in F606W there is an apparent extension and an irregular morphology. Archival optical imaging suggests that the light observed in F606W is dominated by the galaxy, but a transient contribution is possible in the F160W images.}
    \label{fig:hst_img}
\end{figure*}

\grb{} is located in a significant galaxy over-density (Fig.~\ref{fig:hst_img}), which we previously reported in~\citet{levan21_gcn}. In particular, the {\em HST} image overlaps with galaxy cluster 400d J1036$+$5713 at $z=0.203$~\citep{mullis03}. At this redshift, the isotropic equivalent energy of \grb{} would be $E_{\text{iso}}=2\times 10^{51}$\,ergs. The nearest galaxies of the cluster are within a few arcsec of the GRB.

Furthermore, \grb{} is $45$\,kpc ($30$\,arcsec) away from spiral galaxy WISEA J103604.24$+$571327.7, which has an absolute magnitude of $-21.97 \pm 0.50$\,mag, a reported apparent magnitude of $i = 15.962 \pm 0.007$\,mag~\citep{abazajian04}, and a redshift of $z=0.0817$~\citep{abazajian04, albareti17}, where $E_{\text{iso}}=3 \times 10^{50}$\,erg.

Finally, the pre-burst CFHT and Subaru data indicate the presence of a source at the location of \grb{} that is observationally red ($u-r=1.1\pm0.6$). The late-time {\em HST} F606W image (Fig.~\ref{fig:hst_img}) reveals a relatively complex emission region, where the burst is consistent with a compact knot of emission that is marginally extended (major axis is $0.13$\,arcsec), along with a more extended component to the south-east. The extended morphology may be indicative of an underlying galaxy. Table~\ref{tab:photometry_hst} presents {\em HST} photometry using a narrow $0.2$-arcsec aperture centred on the bright knot, as well as a larger $1$-arcsec aperture. The complex nature of the emission is discussed in more detail in Sec.~\ref{sec:lateSN}.

\citet{becerra23} calculate a chance alignment probability of $P_{\text{chance}}=0.02$ for the underlying source, $P_{\text{chance}}=0.05$ for the galaxy at $z=0.0817$, and $P_{\text{chance}}=0.012$ for the cluster at $z=0.203$ (or $10^{-3}$ when factoring in cluster brightness and applying a cut on X-ray flux).

\subsection{Redshift}\label{sec:redshift}

\begin{figure*}
    \centering
    \includegraphics[width=0.9\textwidth]{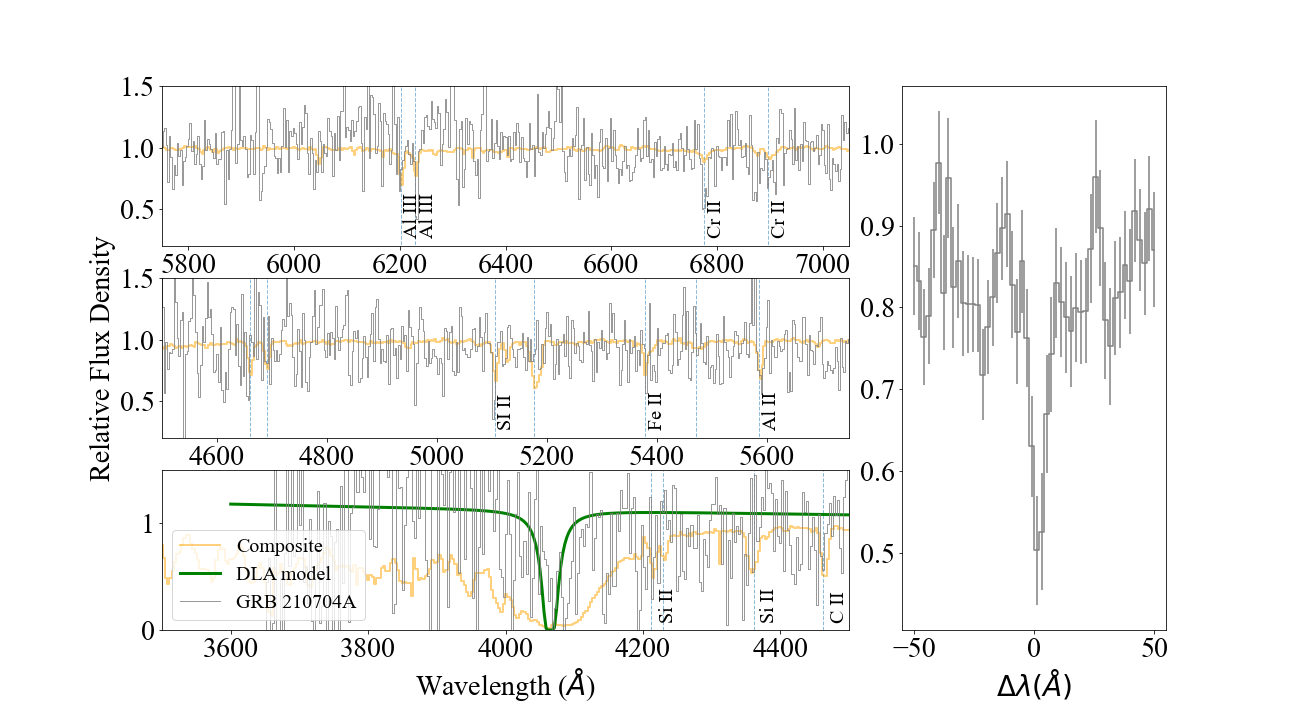}
    \caption{GTC afterglow spectrum of \grb{}. The spectrum is split in wavelength over the three panels on the left. The lower panel includes a damped Ly~$\alpha$ model. The composite spectrum~\protect\citep{christensen11} is also shown. We note that flux calibration is poor below $\approx 4000$\AA. The right panel displays the stack of lines, using a width of $100$\,\AA{} around all the lines labelled in one of the left three panels. Note that the \protect\ion{Al}{ii} line has been omitted from this stack due to a proximate skyline.}
    \label{fig:afterglowspec}
\end{figure*}

In our GTC afterglow spectrum taken at $T_0+1.1$\,d (Fig.~\ref{fig:afterglowspec}), we detect a continuum down to $3800$\,\AA{}, which gives an upper limit on the redshift of $z < 3.17$. The signal-to-noise ratio (S/N) of the afterglow spectrum is low (in the range $3\sigma$--$5\sigma$ per pixel over wavelengths $4000$--$7500$\,\AA{}) and even lower at the blue end ($\lambda < 4000$~\AA), where the flux calibration is also less reliable. We find a broad dip at $\approx 4050$\,\AA{}, which we tentatively interpret as Ly~$\alpha$ absorption. This implies a redshift of $z=2.34$. Low-significance detections of \ion{Si}{ii} and \ion{C}{ii} are consistent with this redshift. \ion{C}{iv} and \ion{Si}{iv} are not detected, although the latter is not unprecedented. We measure for \ion{C}{ii} and \ion{C}{iv} rest-frame equivalent widths of $1.3 \pm 0.4$\,\AA{} and $<1.2$\,\AA{} (3$\sigma$ upper limit), respectively. Comparing these measurement with Fig.~8 of \citet{deugartepostigo12}, we can see that, due to the low S/N, the non-detection of \ion{C}{iv} is not constraining.

To consolidate the identification of the spectral lines, we also performed line-stacking analysis. In particular, we stack the low-ionization metal lines with an equivalent width $>0.5$\,\AA{} based on the line-strength measurements in the composite GRB spectrum of \citet{christensen11}. Due to its proximity to an atmospheric skyline, we omit from the stack \ion{Al}{ii}. The result is shown in the right panel of Fig.~\ref{fig:afterglowspec}. The stacked data indicate a clear absorption feature, corroborating our spectral analysis. Combined with the presence of Ly~$\alpha$ absorption this provides high confidence in the redshift.

The late-time LBT host spectrum reveals weak continuum over the range $4500$--$9200$\,\AA{}. No emission lines are however visible over this range. This is in fact consistent with the absorption redshift, since at $z = 2.34$ none of the brighter lines falls in the covered range, thus providing further support for the GTC value.

\begin{figure}
    \centering
    \includegraphics[width=0.49\textwidth]{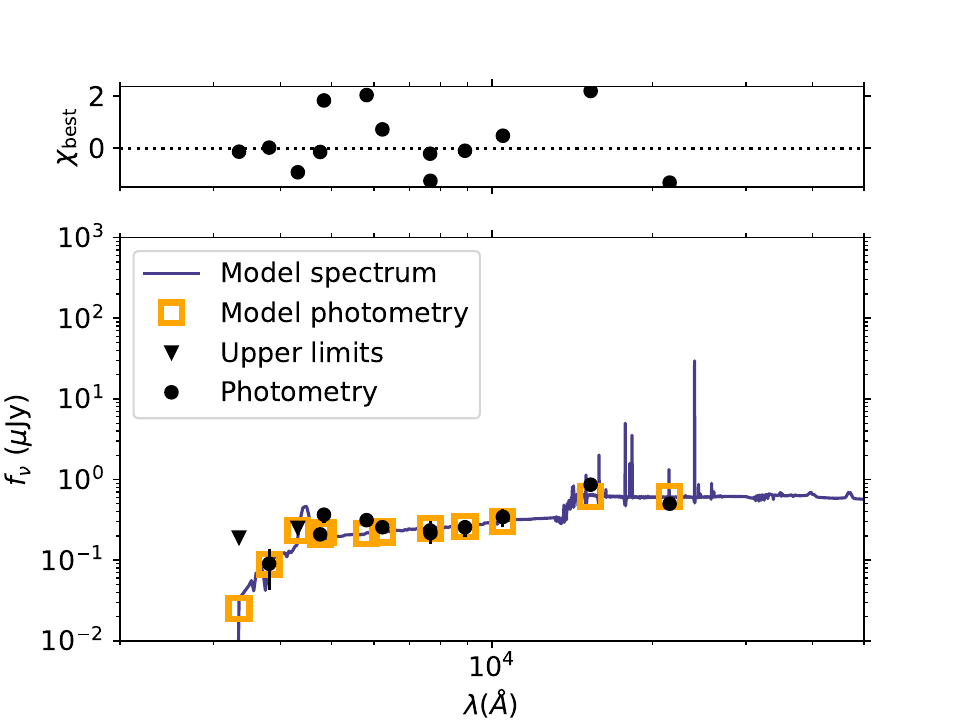}
    \caption{SED fit to the galaxy underlying \grb{}, using $1$-arcsec aperture {\em HST} photometry, the Gemini $K$-band detection at $161$\,d and the archival CFHT and Subaru data. The dotted line in the upper panel corresponds to $\chi_{\text{best}}=0$, the squared mean is $\chi_{\text{best,avg}}^2=0.06$.}
    \label{fig:sedfit}
\end{figure}

To check, we also determine a photometric redshift for the underlying source with {\sc prospector}~\citep{leja17, johnson21}, combining the pre-burst CFHT and Subaru observations, the {\em HST} photometry taken $107$\,d and $194$\,d post-trigger using $1$-arcsec apertures, and the Gemini $K$-band detection from $T_0+161$\,d. We use the delayed-exponential parametric star formation history model, which describes the star formation rate as linearly rising followed by an exponential decline with time. From the SED fit (Fig.~\ref{fig:sedfit}), we derive a redshift of $z=2.63^{+0.21}_{-1.34}$ ($68$\,per cent credible interval). The resulting fit parameters suggest a galaxy of low mass [$\text{log}_{10}\,(M/\text{M}_\odot)\approx9.6$] and low metallicity ($Z\approx0.02\,\text{Z}_\odot$), with a relatively young stellar population ($t\approx0.3$\,Gyr) and moderate extinction $A_V\approx1.0$. This is consistent with the hosts of core-collapse GRB progenitors~\citep{palmerio19}, as is the case for \grb{}'s position with respect to the underlying source, its spectral lag ($\tau=80\pm9$\,ms; \citealt{becerra23}), its $E_{\text{iso}} = 2 \times 10^{53}$\,erg, and its location on the Amati relation at $z=2.34$. This is indicative of a core-collapse progenitor rather than a compact object merger, which is atypical for an EE-GRB.

When we use the more narrow $0.2$-arcsec apertures for the {\em HST} data instead, we obtain consistent SED fit parameters. Setting the redshift to $z=2.34$ in the SED fitting also gives consistent galaxy properties. 

The photometric redshift of the underlying source is consistent with the tentative redshift derived from the afterglow spectrum, thereby reinforcing the credibility of the spectroscopic redshift. We thus consider $z=2.34$ as the redshift of \grb{} for our further analysis.

\subsection{Light Curve: Optical/IR Excess}\label{sec:lc_excess}

\begin{figure}
    \centering
    \includegraphics[width=0.4\textwidth]{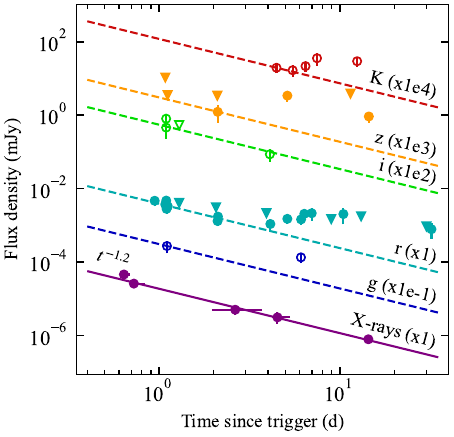}
    \caption{Light curve of \grb{} in different bands in the observer frame. The flux for the visible and IR bands are multiplied with arbitrary powers of ten, as indicated in the brackets on the right, to visually separate them in the figure. Open and closed symbols are alternated per filter for visual clarity. Upper limits are given as triangles. The X-ray data can be fitted with a power law with index $\alpha=-1.2$, shown with the solid line. A power law with the same index is fitted to the pre-excess data of the other bands (dashed lines).}
    \label{fig:lc_afterglow}
\end{figure}

For the light curve analysis in this paper, we correct the photometry in Table~\ref{tab:photometry_optIR} for Galactic extinction $E(B-V)=0.007$~\citep{green15}, following \citet{schlafly11} to obtain the true extinction per band. To account for filter differences and uncertainties in the data reduction, we add an extra systematic error of $0.3$~mag in quadrature to the optical/IR photometry. The light curve of \grb{} is presented in Fig.~\ref{fig:lc_afterglow}. The solid line represents the best-fitting power law for the five X-ray detections, with an index of $\alpha=-1.2\pm0.1$ ($\chi^2_{\text{red}}=0.6$; $F_{\text{X}} \propto t^{\alpha}$). This index is consistent with typical GRB afterglows~\citep{zhang06} and yields an electron index of $p=2.6$ (assuming a uniform medium and the slow-cooling regime). The dashed lines in the figure indicate power laws with the same index fitted to the data of the other bands. Excess emission with respect to the power-law afterglow is evident in all visible and IR bands except $i$, for which we do not have data at the time of the excess. The excess in the $r$ band starts after $4$--$5$\,d, while in the $K$-band it appears to be moderately delayed by $\approx1.5$\,d. The observed peak of the excess is at $T_0+7$\,d ($2$\,d in the rest frame), reaching an absolute magnitude of $M_r=-22.0$\,mag. This is significantly brighter than the pre-burst CFHT data ($\Delta m_r = 2.3\pm0.5$~mag). The upper limit reported by \citet{volnova21_gcn} suggests that the observed peak closely approximates the true peak in $r$. However, due to the limited IR coverage, we can only set limits on the true peak absolute magnitude, $M_K\lid-22.6$\,mag, and peak time, $7$--$12$\,d, in $K$ (rest-frame $r$).

\begin{figure}
    \centering
    \includegraphics[width=0.4\textwidth]{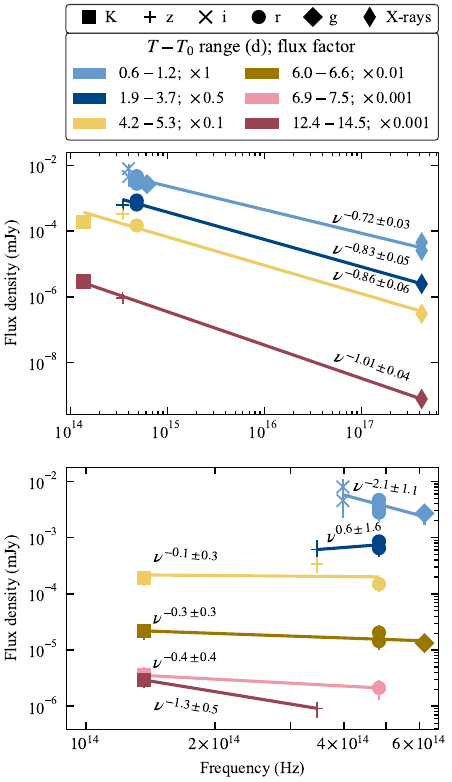}
    \caption{Colour evolution of \grb{}. The upper panel includes X-ray data. The lower panel only contains visible- and IR-band data. In both panels, the time bins are chronologically ordered from top to bottom and visually separated by multiplying the fluxes with the flux factor listed in the legend. A power law ($f\propto \nu^{\beta}$) is fitted to the data of each time bin. The time bins $0.6$--$1.2$\,d and $1.9$--$3.7$\,d post-GRB correspond to the early afterglow phase. The bin of $6.9$--$7.5$\,d is the bin around the peak of the excess emission.}
    \label{fig:colourevolution}
\end{figure}

Fig.~\ref{fig:colourevolution} shows the colour evolution of \grb{}, where the lower panel shows the optical/IR SEDs and the upper panel includes the X-ray detections. Due to the limited number of simultaneous observations, we group the detections in temporal bins (we note that because the behaviour in not monotonic, interpolating to a fixed time is not straightforward). We adopt a default bin width of $0.6$\,d, but extend the second and third epochs to span the full {\em Swift}/XRT observing intervals, yielding widths of $1.8$\,d and $1.1$\,d, respectively. Additionally, we include a wide late-time bin spanning $12.4$--$14.5$\,d that covers the $K$-band detection at $12.4$\,d and the $z$-band and X-ray data at $14.2$--$14.5$\,d. We note that although the inclusion of the $K$-band observation broadens the later time bin, excluding it yields a similar optical/IR to X-ray SED (upper panel of Fig.~\ref{fig:colourevolution}). A power law ($f\propto \nu^{\beta}$) is fitted to the data of each time bin, yielding spectral indices $\beta_{\text{ox}}$ and $\beta_{\text{op}}$ for the upper and lower panel, respectively. 

In the early afterglow phase ($<3.7$\,d), a power-law fit to the X-ray and optical/IR data yields a spectral index of $\beta_{\text{ox}}=-0.75\pm0.03$ (weighted average with $\chi^2_{\text{red}}=3.6$; $f_\nu\propto\nu^{\beta_{\text{ox}}}$). This is consistent with the power-law slopes of the X-ray spectra ($\beta_X=-0.5^{+0.4}_{-0.6}$ and $\beta_X=-0.7^{+1.2}_{-2.3}$ for $0.6$--$1.2$\,d and $1.9$--$3.7$\,d, respectively), although we note the substantial errors in the latter. Moreover, it is consistent with nonthermal synchrotron emission from a standard GRB afterglow~\citep{depasquale03} and the spectral index that can be derived from closure relations: $\beta=2\alpha/3=-0.8$ (assuming a uniform medium and the slow-cooling regime). The optical/IR spectral index $\beta_{\text{op}}=-1.2\pm0.9$~(weighted average, $\chi^2_{\text{red}}=1.8$) is poorly determined, but consistent with $\beta_{\text{ox}}$. This supports a single synchrotron component across the IR to X-ray bands at early times.

The X-ray and optical/IR light curves decouple after $\approx4$\,d (Fig.~\ref{fig:lc_afterglow}). Although no X-ray observations were taken around the light-curve peak, interpolating the X-ray flux density to $7.2$\,d using $\alpha=-1.2$ yields a steeper spectral index of $\beta_{\text{ox}}\approx-1$ at $6.9$--$7.5$\,d. At later times ($12.4$--$14.5$\,d), $\beta_{\text{ox}}$ remains consistent with this steeper index (Fig.~\ref{fig:colourevolution}). The optical/IR spectral index $\beta_{\text{op}}$ remains unchanged during the excess emission phase up to $7.5$\,d, beyond which the large uncertainties limit further interpretation (Fig.~\ref{fig:colourevolution}).

The observed spectral steepening could represent the emergence of a new optical/IR emission component. Alternatively, it could be caused by a cooling break starting to pass through the X-ray band. Although there is no evidence for such a break in the X-ray light curve (Fig.~\ref{fig:lc_afterglow}), the sparse light curve does allow for a cooling break in combination with a refreshed shock, as will be discussed in Section~\ref{sec:refreshedafterglow}.

To further investigate the excess emission, we determine the rise time $t_{1/2,\text{rise}}$ from half-maximum flux density to peak by linearly interpolating the K-band light curve, following \citet{perley20} and \citet{ho23}. Using linear extrapolation, we also determine the fade time $t_{1/2,\text{fade}}$ of the excess emission. The sum of the rise and fade time gives us the duration of the excess above the half-maximum flux. The results are listed in Table~\ref{tab:risefade}. We discuss the duration of the excess further in the next section.

\begin{table}
    \centering
    \caption{Rise and fade times with respect to half-maximum flux for \grb{} and two {\em Einstein Probe} transients, in the observer frame and in the band closest to rest-frame $g$. The sum of the rise and fade time, which is the duration of the transient above half-maximum flux, is listed in the last column (in days).}
    \label{tab:risefade}
    \begin{tabular}{llrrr}
        \hline\hline
        Transient &
        Band &
        $t_{1/2, \text{rise}}$ (d) &
        $t_{1/2, \text{fade}}$ (d) &
        $\Delta t$ (d) \\
        \hline
        GRB~210704A &  $K$ & $1.8 \pm 0.6$ & $14.8\pm 6.4$ & $16.6\pm 6.4$\\
        EP240414a   &  $r$ & $1.2 \pm 0.1$ & $1.3 \pm 0.4$ & $2.5 \pm 0.4$\\ 
        EP241021a   &  $z$ & $2.0 \pm 0.2$ & $5.7 \pm 1.1$ & $7.7 \pm 1.1$\\
        \hline
    \end{tabular}
\end{table}

\section{Discussion}\label{sec:discussion}

\subsection{Comparing \grb{} to Other GRBs}

In principle, the rise of the light curve to a later peak might be the result of viewing a GRB off-axis. However, the rapid rise of the afterglow in the optical/IR excludes this scenario.

\begin{figure}
    \centering
    \includegraphics[width=0.4\textwidth]{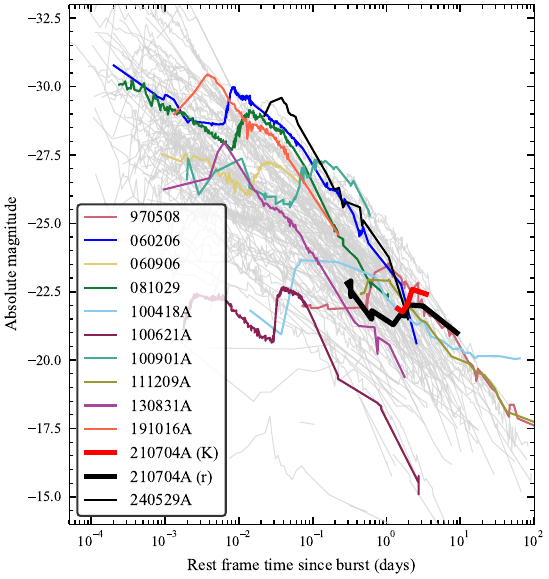}
    \caption{Kann plot comparing the light curve of \grb{} in the $r$-band (rest-frame UV) and in the $K$-band (rest-frame $r$) to the light curves of GRBs. The non-highlighted light curves are from \protect\citet{kann11} in $Rc$, just like the light curves of GRBs 970508, 060206, 060906, 081029, 100621A, and 100901A which are highlighted in colour as they undergo rebrightening episodes. Other rebrightening bursts that are included are GRBs 100418A~($r'/R$; \protect\citealt{deugartepostigo18}), 111209A~($r'$; \protect\citealt{kann18}), 130831A~($V$; \protect\citealt{depasquale16}), 191016A~($r$; \protect\citealt{pereyra22}), and 240529A~($r/R$; \protect\citealt{sun24}). Light curves are not corrected for host galaxy extinction.}
    \label{fig:kann}
\end{figure}

The Kann plot in Fig.~\ref{fig:kann} shows that \grb{} starts off as an average afterglow, but due to the excess emission it become as bright as the brightest afterglows observed. The excess corresponds to an observed luminosity increase of $0.62\pm0.24$\,mag in $K$ and $0.69\pm0.40$\,mag in $r$. This is both consistent with a rebrightening as well as with a flattening (within $3\sigma$) of the light curve. Late-time flattening in GRB light curves is typically due to host emission, but that is not the case here as the CFHT data reveal a much fainter underlying galaxy ($\Delta m_r = 2.3\pm0.5$~mag). Late-time light curve plateaus can also be driven by the same physical mechanisms responsible for rebrightenings, but with a reduced energy input. For instance, refreshed shocks have been proposed to explain the plateau phase of GRB~191016A around $0.02$\,rest-frame days~(Fig.~\ref{fig:kann}; \citealt{pereyra22}). 

The coloured graphs in Fig.~\ref{fig:kann} show GRBs that undergo rebrightening episodes. The rebrightening typically occurs during the first rest-frame day and corresponds to a luminosity increase of $1$--$3$\,mag. GRBs with multi-band data show that the colour of the emission is bluer before the rebrightening~(e.g. GRBs 081029, 100621A, and 111209A; \citealt{deugartepostigo18}). While the observed light curve of \grb{} could be consistent with a rebrightening and reddening, the photometric uncertainties preclude a definitive match with these properties of rebrightening GRBs. \grb{}'s excess emission also rises later than the rebrightening GRBs in Fig.~\ref{fig:kann}. The physical origin of rebrightenings in GRBs remains debated, but proposed models for the highlighted rebrightening GRBs in Fig.~\ref{fig:kann} include refreshed shocks (e.g. GRB~100621A; \citealt{greiner13}) or a CSM density bump (e.g. GRB~970508; \citealt{tam05}). Section~\ref{sec:redback} further examines these models in the context of \grb{}.

\subsection{Comparing \grb{}'s Excess Emission to Other Transient Populations}

\begin{figure}
    \centering
    \includegraphics[width=0.48\textwidth]{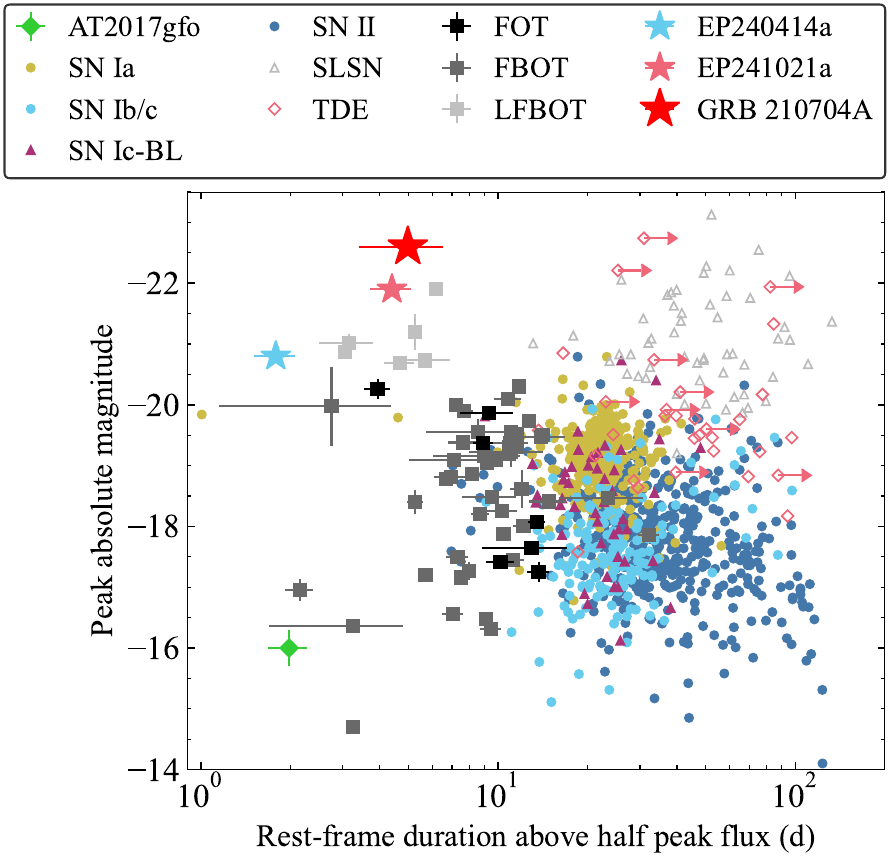}
    \caption{Peak absolute magnitude vs. the rest-frame duration above half the peak flux, using the observed peak. We include kilonova AT2017gfo (rest-frame $g$; \protect\citealt{villar17}), SNe (rest-frame $g/r$; \protect\citealt{perley20}), SLSNe \& TDEs (rest-frame $u/g/r$; \protect\citealt{perley20}), FOTs (rest-frame $g/r$; \protect\citealt{ho23}), and FBOTs \& LFBOTs (band closest to rest-frame $g$; \protect\citealt{ho23}; \protect\citealt{pursiainen25}). The LFBOTs are empirically selected as FBOTs with $M_{\text{peak}}<-20.5$\,mag. We also plot EP240414a (rest-frame $g$), EP241021a (rest-frame $g$), and \grb{} (rest-frame $r$), for which we calculated the duration (Table~\ref{tab:risefade}). Magnitudes were corrected for Galactic extinction and in the case of iPTF15ul also for host extinction~\protect\citep{perley20}.}
    \label{fig:peakcomparison}
\end{figure}

A light curve rise and peak can also be caused by transient light, including that from kilonovae, SNe (Type Ia, Ib/c, Ic-BL, and II), superluminous SNe (SLSNe), tidal disruption events (TDEs), fast optical transients (FOTs), and the empirical subpopulations of fast blue optical transients (FBOTs; where $g-r\loa-0.2$ following \citealt{ho23}) and luminous FBOTs (LFBOTs; which we define as FBOTs with $M_{\text{peak}}<-20.5$ in rest-frame $g$). We compare \grb{} to these transients in Fig.~\ref{fig:peakcomparison}, where we plot the peak absolute magnitude versus the duration above half-maximum flux in the rest frame. 

The excess emission in \grb{} is much more luminous than prototypical kilonova AT2017gfo and can therefore not be explained by the typical SGRB progenitor scenario. A brighter kilonova may be possible if it is powered by a magnetar. We discuss this scenario further in Sec.~\ref{sec:alternative_models}, where we perform modelling.

\begin{figure*}
    \centering
    \includegraphics[width=\textwidth]{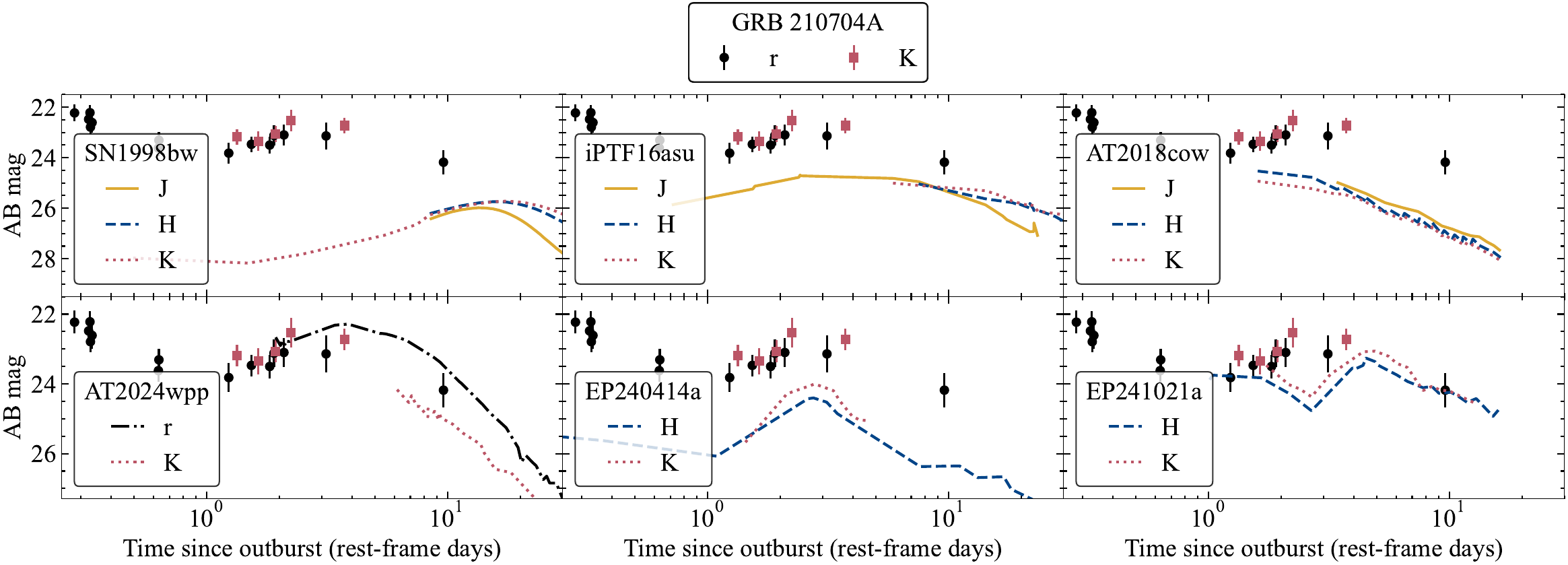}
    \caption{Light curves in the observer frame, comparing \grb{} to prototypical LGRB with SN Ic-BL: GRB~980425 / SN1998bw~\protect\citep{patat01}, FBOT iPTF16asu which is associated to an SN Ic-BL~\protect\citep{whitesides17}, prototypical LFBOT AT2018cow~\protect\citep{prentice18}, the brightest LFBOT to date: AT2024wpp~\protect\citep{lebaron26}, EP240414a~\protect\citep{vandalen25}, and EP241021a~\protect\citep{busmann25, quirolavasquez26}. All transients were converted to the $z=2.34$ frame.}
    \label{fig:lc_comparison}
\end{figure*}

The luminosity and duration of \grb{}'s excess are also inconsistent with a typical SN Ic-BL that is commonly associated with LGRBs (Fig.~\ref{fig:peakcomparison}). Additionally, \grb{}'s excess rises on a shorter time-scale than SNe Ic-BL (upper left panel of Fig.~\ref{fig:lc_comparison}). We can exclude other standard core-collapse SN (Type Ib/Ic/II) or thermonuclear (Ia) SN progenitors for the same brightness and temporal reasons (Fig~\ref{fig:peakcomparison}). SLSNe and TDEs match \grb{} in brightness, but evolve on longer time-scales.

\grb{}'s excess emission is more luminous and more rapid than most FBOTs (Fig.~\ref{fig:peakcomparison} and second panel of Fig.~\ref{fig:lc_comparison}). It is also brighter and redder than prototypical LFBOT AT2018cow (upper right panel of Fig.~\ref{fig:lc_comparison}). However, we note that the colour may not be as important given the similarities between nonblue FOT AT2020bot (upper left black square in Fig.~\ref{fig:peakcomparison}) and the blue LFBOT population. The GRB's emission is on the bright end of the population of LFBOTs (Fig.~\ref{fig:peakcomparison}), comparable in brightness to most luminous LFBOT known to date -- AT2024wpp (lower left panel of Fig.~\ref{fig:lc_comparison}). 

Luminous and rapidly brightening optical counterparts have been observed for some fast X-ray transients (FXTs). The {\em Einstein Probe} FXT EP240414a~\citep{bright25, srivastav25, sun25, vandalen25} is a prime example. Its light curve consists of an early decay, a rebrightening episode that peaks at $2.9$\,d in the rest frame ($z=0.401$), and a late-time fainter rebrightening that starts after $7$~rest-frame days (Fig.~\ref{fig:lc_comparison}). The brightest ($2.9$\,d) peak resembles the excess in \grb{} in peak time and colour (fifth panel of Fig.~\ref{fig:lc_comparison}), although it differs in peak magnitude and in peak sharpness (Fig.~\ref{fig:peakcomparison}). EP240414a has a massive star origin, as late-time spectroscopy reveals an SN Ic-BL~\citep{vandalen25}. The radio emission strongly resembles a moderately relativistic afterglow typical for collapsar LGRBs~\citep{bright25}, despite no gamma-ray counterpart having been detected. A possible gamma-ray counterpart of EP240414a is constrained to $L_{\text{iso}}<10^{51}$\,erg\,s$^{-1}$~\citep{bright25}, which could suggest that EP240414a is a low-luminosity GRB event. This is in stark contrast to \grb{}, for which the LAT detection indicates a high bulk Lorentz factor. Motivated by EP240414a's soft X-ray spectrum (peaking $<1.3$\,keV), its spectral features, and its large offset from its peculiar host galaxy, \citet{sun25} propose a stripped-envelope Wolf-Rayet star progenitor that launches a weaker jet than typical for LGRB progenitors, potentially due to a smaller core angular momentum. For the main optical peak in EP240414a's light curve, \citet{srivastav25} suggest a refreshed shock scenario. However, just like for \grb{}, the X-ray light curve of EP240414a declines and the spectral index $\beta_{\text{ox}}$ steepens during the main optical peak. \citet{vandalen25} argue that this decoupling of the X-ray and optical light curves disfavours an afterglow origin. They propose a low-luminosity collapsar GRB scenario with early cocoon emission, a bright peak due to SN-CSM interaction, and the late peak corresponding to SN emission. We explore both refreshed shocks and SN-CSM interaction as possible origins of \grb{}'s excess in Sec.~\ref{sec:redback}. 

Another optically luminous and rapid FXT is EP241021a~\citep{busmann25, gianfagna25, shu25, wu25, yadav25, quirolavasquez26}. It is not detected in gamma rays, but the limits on the gamma-ray emission are shallow~\citep{busmann25}. EP241021a's optical light curve exhibits an initial afterglow-like decay, followed by a rapid rebrightening peak at $4.4$~rest-frame days ($z=0.748$) and a shallower peak at $11$~rest-frame days (lower right panel of Fig.~\ref{fig:lc_comparison}). The late-time peak can be explained by an emerging SN, implying a collapsar origin for EP241021a~\citep{gianfagna25, quirolavasquez26}. The radio afterglow points to an at least mildly relativistic outflow, connecting EP241021a to GRB-like explosions or relativistic TDEs~\citep{yadav25}. The brightest peak of EP241021a resembles the peak of \grb{} in colour, magnitude jump, and peak magnitude (Fig.~\ref{fig:lc_comparison}) as well as in peak duration (Fig.~\ref{fig:peakcomparison}), although \grb{} peaks earlier. The X-ray light curve also behaves differently, with that of EP241021a tracking the optical rebrightening. \citet{busmann25} favour the refreshed shock scenario for the non-thermal brightest peak in EP241021a. However, \citet{quirolavasquez26} argue that the rise time of the peak is shorter than expected for a refreshed shock and that the fast rebrightening is extremely difficult to reconcile with any interpretation. \citet{shu25} agree that the brightest peak in EP241021a cannot be well explained by existing models, including SN shock breakout, a merger-triggered magnetar, a highly structured jet or a repeating TDE.

In any case, the empirical similarity of \grb{} to these recent {\em Einstein Probe} transients is notable, in particular since those systems were both much closer than \grb{}, but had no associated $\gamma$-ray emission. While similarity in emission does not necessarily imply a causal connection, it is none the less interesting to consider this difference should such a connection exist. In particular, the powerful jet in FXTs could be absent because it is not produced (e.g. because of jet launch conditions), because it is choked by heavy baryon loading in the progenitor star, or because of viewing angle effects.

\subsection{Possible Late-time Supernova}\label{sec:lateSN}

As stated in Sec.~\ref{sec:environment}, the morphology of the underlying source in the late-time {\em HST} data is characterised by a compact emission knot on top of an extended component (right panel of Fig.~\ref{fig:hst_img}). An investigation of the emission colour reveals a red source in both {\em HST} epochs that yields an even redder colour for smaller apertures (from $\text{F105W}-\text{F160W}=1.0\pm0.1$ in a $0.42$-arcsec ($6$-pixel) aperture to $\text{F105W}-\text{F160W}=1.5\pm0.1$ in a $0.14$-arcsec aperture). This is unlikely to be caused by extinction, as $\text{F606W}-\text{F105W}$ is blue. In the absence of an exceptionally strong Balmer line, this could indicate the presence of a red transient on top of a galaxy. Assuming $z=2.34$, the {\em HST} epochs were taken $32$\,d and $58$\,d after the GRB in the rest frame. This is late, but not unheard of, for an SN associated with a GRB to be detected~\citep{woosley21}. 

\begin{figure}
    \centering
    \includegraphics[width=0.49\textwidth]{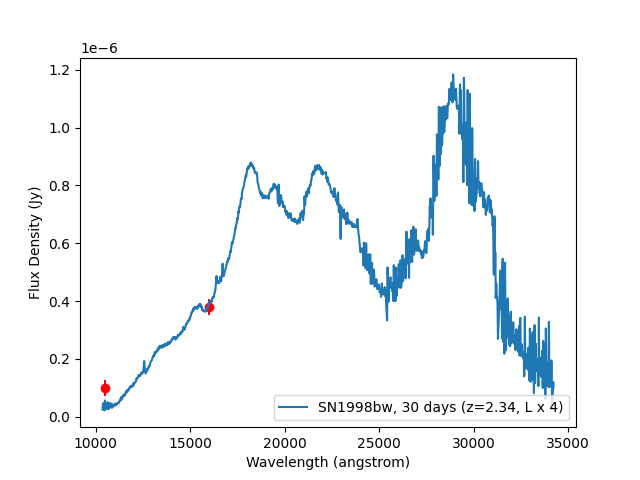}
    \caption{The spectrum of SN1998bw -- a prototypical SN Ic-BL -- (graph) taken $30$~rest-frame days after its associated GRB~\protect\citep{galama98,patat01}, redshifted using $z=2.34$ and shifted to four times its luminosity to roughly match the {\em HST} IR detections of \grb{} (data points).}
    \label{fig:SNfit}
\end{figure}

Fig.~\ref{fig:SNfit} compares the IR {\em HST} detections to a spectrum of a typical Type Ic-BL SN associated with a GRB: SN1998bw. The colour of the {\em HST} source is consistent with that of an SN, but the brightness of SN1998bw must be increased by a factor of four to align with our data. Accordingly, the transient would have an absolute magnitude of $M_B=-20$\,mag, at $30$\,d after the GRB in its rest frame. That is brighter than what is typically observed in Type Ic SNe (Fig.~\ref{fig:peakcomparison}; \citealt{kasliwal12}). However, we note that the underlying galaxy likely contributes to the total brightness. We further investigate the SN scenario in our modelling (Sec.~\ref{sec:redback}).

GRB~241105A (\citealt{dimple25}; $z=2.681$) shares several characteristics with \grb{}. It exhibits a bright, short initial spike followed by weaker emission up to $64$\,s. Although no accompanying SN was detected, {\em James Webb Space Telescope} observations reveal host-galaxy properties that are more typical of the LGRB population (active star formation and a low metallicity). Combined with the small sub-kpc offset from the host and afterglow modelling that requires a high-density CSM, a collapsar origin for GRB~241105A is plausible. However, we note that the prolonged emission in GRB~241105A is spectrally harder than its initial spike, which is atypical for EE-GRBs (e.g. \citealt{kaneko15}). In this respect, \grb{} more closely resembles the canonical EE-GRB population. Detecting an SN associated with \grb{} would challenge the standard picture where EE-GRBs originating solely from compact object mergers. New {\em HST} imaging would be able to confirm whether an SN signature was indeed present in the late-time epochs.

\subsection{Light Curve Modelling with {\sc redback}}\label{sec:redback}

We have shown that the excess emission of \grb{} is not easily matched to the emission observed in other transients, although substantial similarities can be found to some classes. In this section, we attempt to model the light curve of \grb{} with combined afterglow, interaction and/or transient models. We adopt a joint fitting approach, rather than fitting individual model components or parameters independently, in order to avoid introducing significant biases in the inferred posterior distributions~\citep{wallace25}. For the joint fitting, we use Bayesian interference software package {\sc redback}~\citep{sarin24, sarin25} and the {\sc nessai} sampler~\citep{williams21, williams23, williams25} wrapped with {\sc bilby}~\citep{ashton19, talbot25}. We use a standard Gaussian likelihood and the default model priors in {\sc redback}, except for the redshift which we set to $z=2.34$. We include the late-time {\em HST} and Gemini photometry in the modelling, as the late-time data may include an SN component (Sec.~\ref{sec:lateSN}).

\subsubsection{SN-CSM Interaction}\label{sec:csmmodel}

\begin{figure}
    \centering
    \includegraphics[width=0.49\textwidth]{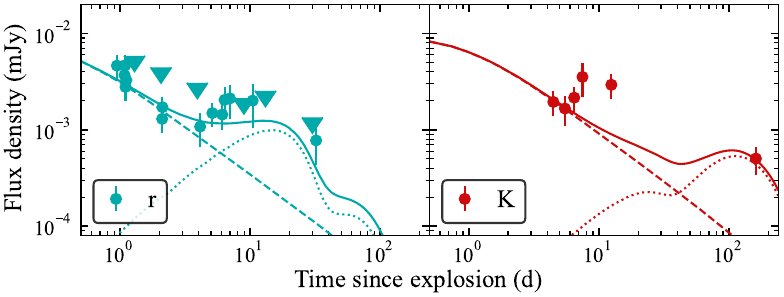}
    \caption{Top-hat afterglow model~\protect\citep{ryan20} combined with an SN-CSM interaction model~\protect\citep{margalit22} and an SN model~\protect\citep{arnett82}, fitted to the $r$- (left panel) and $K$-band (right panel) data of \grb{} (solid graph). The dashed and dotted graphs represent the afterglow model and the combined CSM shock and SN model, respectively. The SN-CSM interaction and SN models share a common photosphere and diffusion description and are therefore plotted as one graph, where the early-time peak is due to the CSM shock and the late-time peak is due to the SN. The model fits correspond to the maximum posterior (prior\,$\times$\,likelihood).}
    \label{fig:redbackfit_csm_split}
\end{figure}

\begin{figure*}
    \centering
    \includegraphics[width=0.7\textwidth]{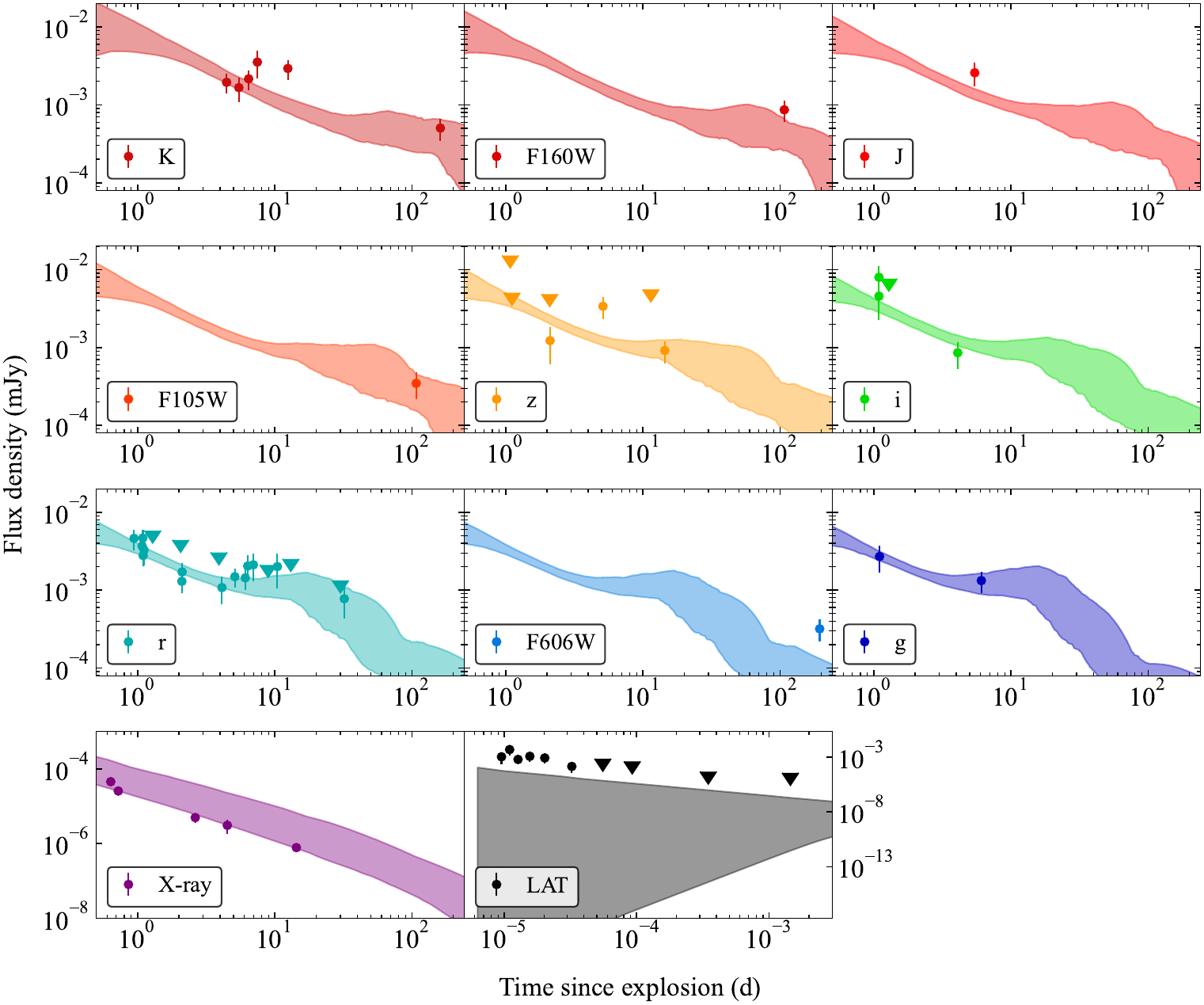}
    \caption{Top-hat afterglow model~\protect\citep{ryan20} combined with an SN-CSM interaction model~\protect\citep{margalit22} and an SN model~\protect\citep{arnett82}, fitted to the light curve of \grb{} with {\sc redback} using 1000 live points. The afterglow model includes inverse Compton scattering and jet spreading. The shaded areas are the $90$\,per cent credible intervals.}
    \label{fig:redbackfit_csm}
\end{figure*}

Motivated by the similarities with EP240414a, we first use a combined model of a top-hat afterglow, CSM shock breakout, and an SN. In particular, we use the semi-analytical top-hat model from {\sc afterglowpy}~\citep{ryan20}. This model assumes a decelerating jet in a homogeneous medium and approximates the jet as a single-shell~\citep{vaneerten10, vaneerten18} to compute light curves. Relativistic beaming and (conical) jet spreading are taken into account. We also include inverse Compton scattering in an attempt to simultaneously explain the high-energy LAT emission. For the CSM interaction, we use the spherically-symmetric one-zone model from \citet{margalit22}. The model assumes a one-zone shell of ejecta, interacting with a CSM shell of thickness $\Delta R_{\text{shell}}$ and mass $M_{\text{CSM}}$ with a constant density (top-hat) profile, located at distance $R_{\text{shell}}$ from the progenitor. The CSM shell is assumed to expand homologously, starting at the time of shock breakout (when photons are first able to escape the shocked CSM). The model explicitly solves the radiative diffusion equation to capture energy transport and losses. To model the SN emission produced by the radioactive decay of $^{56}\text{Ni}$, we use the well-established analytical framework of \citet{arnett82}. We note that the models we use are highly simplified, e.g. assuming spherical symmetry and a single zone. Nonetheless, these models have proven to be successful in reproducing observed light curve features (e.g. \citealt{vandalen25}).

The combined model fit is shown in Fig.~\ref{fig:redbackfit_csm_split} and Fig.~\ref{fig:redbackfit_csm}, where the former highlights the contributions of the different models and the latter shows the 90~per cent credible intervals of the combined model fitted to all available data. The combined model fails to explain all the data. Although the excess emission in $r$ could be consistent with SN-CSM interaction, the model cannot account for the contemporaneous excess in $K$ or $z$. We also note that the afterglow model, despite including inverse Compton scattering, struggles to explain the LAT emission of \grb{}.

The posterior distribution is shown in Fig.~\ref{fig:corner_csm}. The afterglow posterior is not well constrained, but generally implies a highly energetic jet with a relatively large core angle, that is observed on-axis. The jet angle may be overestimated as the modelling assumes a top-hat jet rather than a structured jet. The initial Lorentz factor is unconstrained by the modelling, which is expected as there are no early-time X-ray/optical/IR observations. We note that the CSM interaction model produces a high ejecta velocity and high CSM mass, which is in stark contrast to the prior distributions which favour smaller values. The CSM parameters are likely overestimated as a result of the simplifications made in the modelling, such as the one-zone assumption. The same is true for the unusually high nickel fraction, mass and velocity found by the simple one-zone SN model. Including additional physics could result in better parameter estimates. For example, including nickel mixing could yield a lower nickel fraction and ejecta velocity. Although the modelling produces a low opacity to gamma rays and a low temperature floor, this is in line with the prior distributions and therefore not unexpected.

\subsubsection{Refreshed Afterglow}\label{sec:refreshedafterglow}

\begin{figure*}
    \centering
    \includegraphics[width=0.7\textwidth]{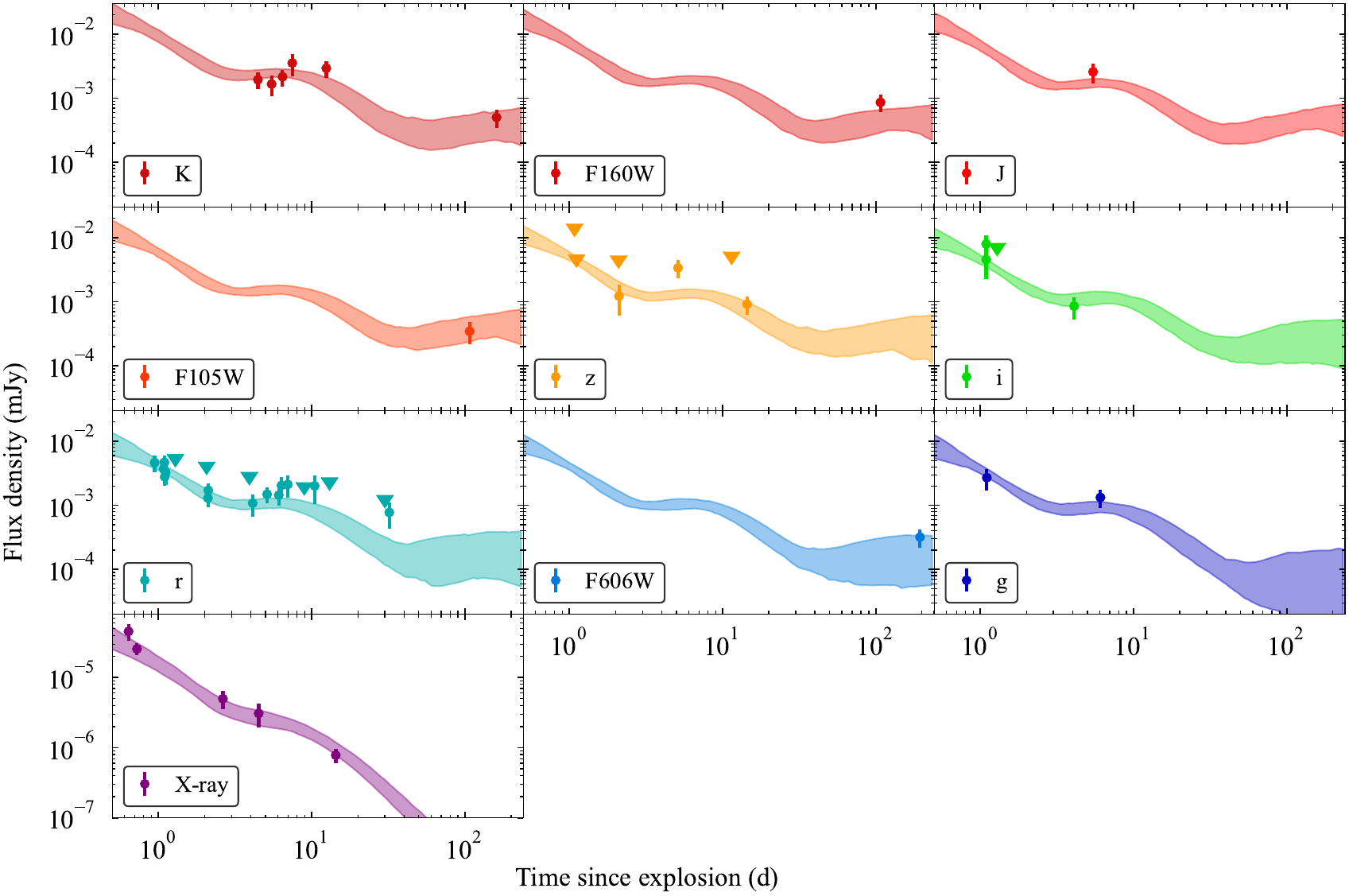}
    \caption{Refreshed top-hat afterglow model~\protect\citep{lamb19, lamb20} combined with an SN model~\protect\citep{arnett82}, fitted to the light curve of \grb{} with {\sc redback} using 500 live points. The shaded areas are the $90$\,per cent credible intervals.}
    \label{fig:redbackfit_refreshed}
\end{figure*}

\begin{figure}
    \centering
    \includegraphics[width=0.49\textwidth]{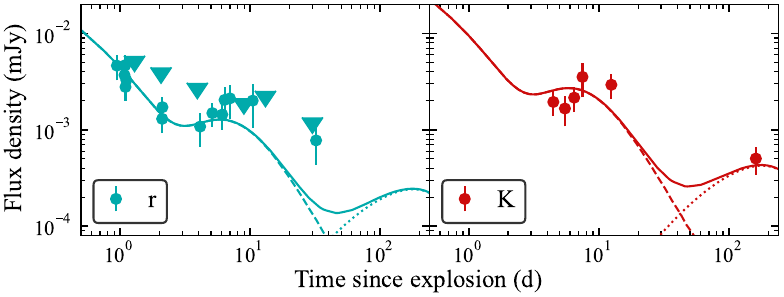}
    \caption{Refreshed top-hat afterglow model~\protect\citep{lamb19, lamb20} combined with an SN model~\protect\citep{arnett82}, fitted to the $r$- (left panel) and $K$-band (right panel) data of \grb{} (solid graph). The dashed and dotted graphs represent the refreshed afterglow and the SN model, respectively. The model fits correspond to the maximum posterior (prior\,$\times$\,likelihood).}
    \label{fig:redbackfit_refreshed_split}
\end{figure}

The refreshed shock afterglow model was proposed for EP240414a~\citep{srivastav25} and EP241021a~\citep{busmann25}, both of which resemble \grb{} in many ways. We therefore investigate the refreshed shock scenario in more detail. For hydrodynamically accelerated GRB ejecta, \citet{ioka05} calculate that refreshed shocks can only produce bumps with $t_{1/2,\text{rise}}/t_{\text{peak}}\gid0.25$. For \grb{}, the observed ratio is $\approx 0.4\pm0.2$ for $r$ and $\approx 0.2 \pm 0.1$ for $K$, making the refreshed shock scenario a possibility. We fit a combined model of a refreshed top-hat afterglow and an SN to the light curve of \grb{}. 

For the SN model, we again use the analytical model from \citet{arnett82}. For the refreshed top-hat afterglow, we use the model from \citet{lamb19, lamb20}, see also \citet{sarin24}, where the model is included in {\sc redback}. This model assumes that the refreshed afterglow is the result of energy injection due to a collision between a slower catching shell and the decelerating impulsive shell. The resultant merged shell is the sum of the two shell masses. The energy injection is modelled as a mild collision, i.e. no secondary shock is produced within the hot impulsive shell at collision and the two shells stick together (see \citealt{zhang02} for details). The afterglow emission model includes synchrotron self-absorption and pre-deceleration physics, where the jet's coasting phase is included before the deceleration time. The model includes a jet spreading description that follows either of the two cases in \citet{granot12} ($a=1$ case is used). However, as this model does not include inverse Compton scattering, we exclude the LAT data from the modelling here.

The combined model fit is shown in Fig.~\ref{fig:redbackfit_refreshed} and Fig.~\ref{fig:redbackfit_refreshed_split}, where the former shows the 90~per cent credible intervals of the combined model fitted to all available data and the latter highlights the contributions of the different models. While the X-ray data do not require a refreshed shock, the refreshed-afterglow model provides an adequate description of them (Fig.~\ref{fig:redbackfit_refreshed}). The late-time steepening of the X-ray light curve is consistent with a cooling break passing through the X-ray band which, following the closure relations, would result in a slope of $\alpha=-1.5$ (using electron index $p=2.6$ as determined earlier and assuming a uniform medium and the slow-cooling regime). 

The posterior distribution is shown in Fig.~\ref{fig:corner_refreshed}. Similar to what was discussed in Sec.~\ref{sec:csmmodel}, the afterglow modelling is unable to constrain the initial Lorentz factor and produces a large core angle, which are likely the result of not having early-time observations and of the top-hat assumption in the modelling, respectively. We note that the light curve of \grb{} can only be explained by a strong refreshed shock (Fig.~\ref{fig:corner_refreshed}): it requires a dense CSM (number density $n_{\text{ism}}\approx50\,\text{cm}^{-3}$), a large Lorentz factor of the shell at the onset of energy injection $\Gamma_1\approx4.8$, a kinetic energy increase by a factor of $\approx18$, a relatively high energy injection index ($s\approx9$ and consistent with a uniform Lorentz factor throughout the catching shell), and a high electron energy fraction $\epsilon_{\text{e}}\approx0.5$. The model thus requires the trailing ejecta shell, which catches up with the earlier ejected shell to cause the refreshed shock, to be highly energetic. Such a strong shock can explain the optical/IR excess emission around $\approx5$--$15$\,d reasonably well, although the model struggles to explain the $z$-band peak (Fig.~\ref{fig:redbackfit_refreshed}). 

Additionally, the modelling shows that most of the late-time data can be decently explained as SN emission. The combined model struggles with the $r$-band detection at $32$\,d. However, the SN model used here is a simple one-zone model~\citep{arnett82} that does not include nickel mixing. When $^{56}\text{Ni}$ is mixed with the outer layers of the SN ejecta, the energy produced by the radioactive decay of $^{56}\text{Ni}$ dissipates more easily through the ejecta, resulting in earlier SN emission~(e.g. \citealt{bersten12, rastinejad25}). An earlier rise in the SN light curve could explain \grb{}'s $r$-band detection at $32$\,d, as well as better explain the detection in F160W (Fig.~\ref{fig:redbackfit_refreshed}). The elevated nickel fraction, ejecta mass and ejecta velocity are again likely an artefact of the simple SN model, where e.g. nickel mixing is excluded and one zone is assumed.

\subsubsection{Shock Cooling}

\begin{figure}
    \centering
    \includegraphics[width=0.49\textwidth]{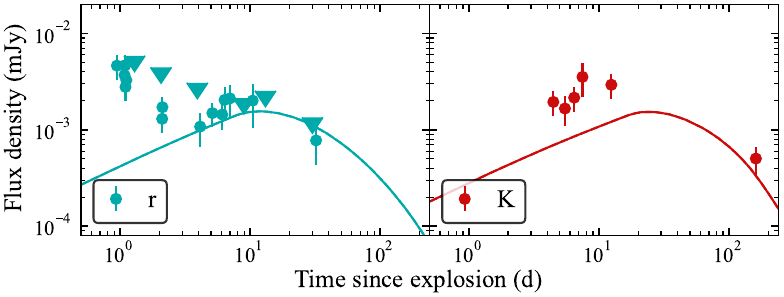}
    \caption{Top-hat afterglow model~\protect\citep{ryan20} combined with a shock cooling model~\protect\citep{piro21} and an SN model~\protect\citep{arnett82}, fitted to the $r$- (left panel) and $K$-band (right panel) data of \grb{} (solid graph). The model fit correspond to the maximum posterior (prior\,$\times$\,likelihood). The shock cooling and SN contributions are negligible in the maximum posterior fit.}
    \label{fig:redbackfit_shockcooling_split}
\end{figure}

\begin{figure*}
    \centering
    \includegraphics[width=0.7\textwidth]{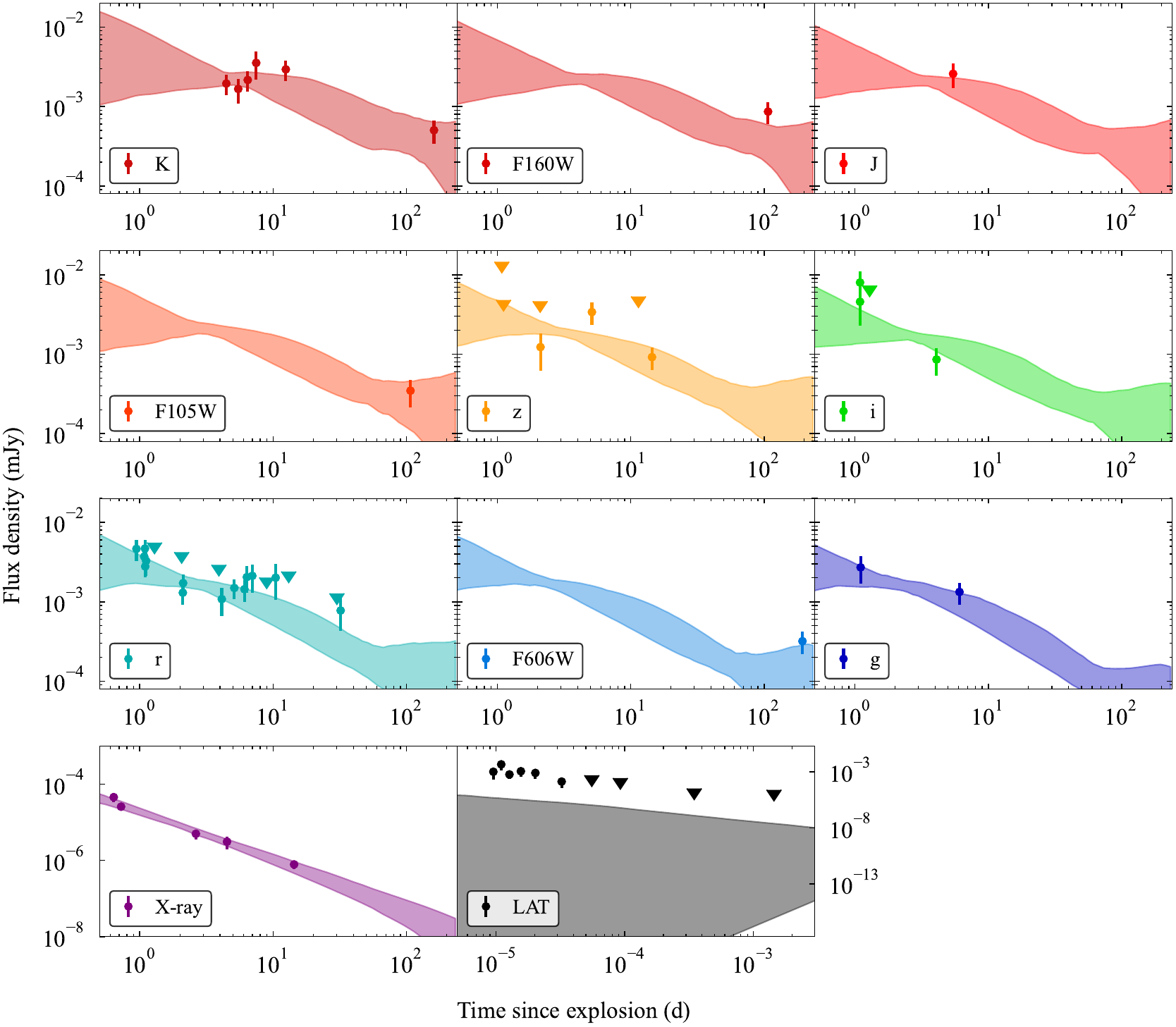}
    \caption{Top-hat afterglow model~\protect\citep{ryan20} combined with a shock cooling model~\protect\citep{piro21} and an SN model~\protect\citep{arnett82}, fitted to the light curve of \grb{} with {\sc redback} using 1000 live points. The shaded areas are the $90$\,per cent credible intervals.}
    \label{fig:redbackfit_shockcooling}
\end{figure*}

We also attempt to fit \grb{}'s light curve with a combined top-hat afterglow, shock cooling, and SN model. As in Section~\ref{sec:csmmodel}, we use the top-hat afterglow model from {\sc afterglowpy}~\citep{ryan20}, including jet spreading and inverse Compton scattering. For the SN model, we use the analytical model from \citet{arnett82} as before. For the shock cooling emission, we use the analytical model from \citet{piro21}. In this model, a shock wave deposits energy $E_{\text{en}}$ into extended material of mass $m_{\text{en}}$ and radius $R_{\text{en}}$, causing the material to expand homologously. The model only considers the homologously expanding phase and assumes a constant electron scattering opacity (motivated by the hot ejecta temperatures). The expanding ejecta are considered to have a two-component profile~\citep{chevalier89}, where the outer component has a strong velocity gradient and the inner ejecta have a more modest velocity gradient. The power-law slopes of the density profiles are parameterised by $nn$ and $\Delta$ for the outer and inner ejecta, respectively. This is a common assumption for density profiles in all models involving a massive star progenitor.

The combined model fit is shown in Fig.~\ref{fig:redbackfit_shockcooling_split} and Fig.~\ref{fig:redbackfit_shockcooling}, where the former highlights the contributions of the different models and the latter shows the 90~per cent credible intervals of the combined model fitted to all available data. Although the data largely falls in the $90$\,per cent credible interval of the combined model, the maximum posterior curve is a poor fit to the data (Fig.~\ref{fig:redbackfit_shockcooling_split}).

The posterior distribution is shown in Fig.~\ref{fig:corner_shockcooling}. As in our previous modelling, the large jet core angle and nickel fraction are likely overestimated due to the model assumptions and missing physics. We note that the posterior of the shock cooling model is very poorly constrained. Combined with the poor maximum posterior fit, we therefore disfavour the combined top-hat afterglow, shock cooling and SN model.

\subsubsection{Alternative Models}\label{sec:alternative_models}

\begin{figure}
    \centering
    \includegraphics[width=0.49\textwidth]{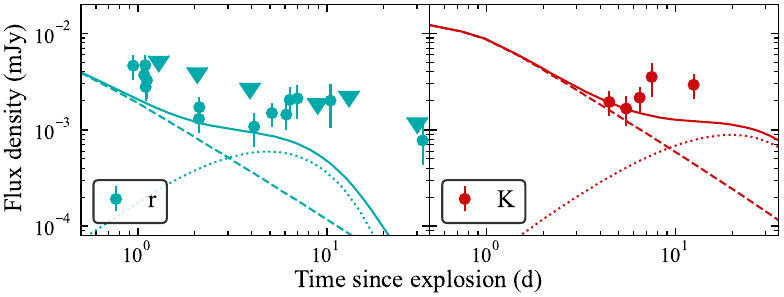}
    \caption{Top-hat afterglow model~\protect\citep{lamb18, sarin24} combined with a general magnetar-driven kilonova model~\protect\citep{sarin22}, fitted to the $r$- (left panel) and $K$-band (right panel) data of \grb{} (solid graph). The dashed and dotted graphs represent the afterglow and the magnetar-driven kilonova model, respectively. The model fits correspond to the maximum posterior (prior\,$\times$\,likelihood).}
    \label{fig:redbackfit_magdrivenKN_split}
\end{figure}

\begin{figure*}
    \centering
    \includegraphics[width=0.7\textwidth]{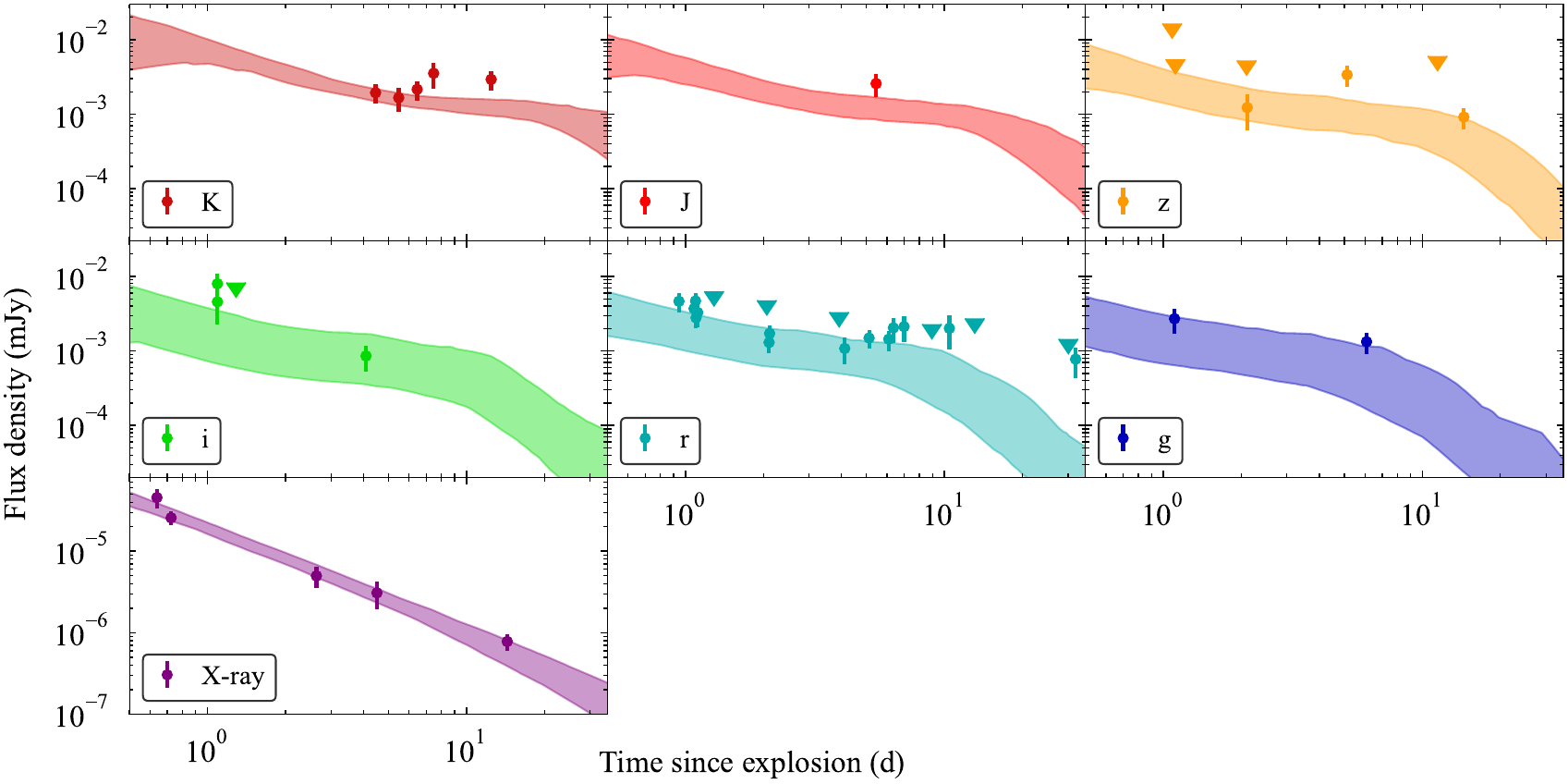}
    \caption{Top-hat afterglow model~\protect\citep{lamb18, sarin24} combined with a general magnetar-driven kilonova model~\protect\citep{sarin22}, fitted to the light curve of \grb{}. The shaded areas are the $90$\,per cent credible intervals.}
    \label{fig:redbackfit_magdrivenKN}
\end{figure*}

Alternative models for light curve bumps include the patchy shell model~\citep{meszaros98, kumar00}, a density bump in the CSM~\citep{wang00, dai02, lazzati02, nakar03} or a magnetar-driven transient. The patchy shell model describes a jet where the energy varies randomly across its head, resulting in 'hot spots' or 'sub-jets'. The excess in \grb{} violates the model requirement of $t_{1/2,\text{rise}}>t_{\text{peak}}$~\citep{ioka05}. The density bump scenario is also implausible, since \grb{} violates the possible maximum amplitude of the rebrightening $\Delta F_\nu/F_\nu\lid1.6\,t_{1/2,\text{rise}}/t_{\text{peak}}$ assuming a typical on-axis afterglow~\citep{ioka05}, with $\Delta F_\nu/F_\nu=0.7$ for \grb{} in K (rest-frame $r$). To investigate a possible magnetar-driven kilonova origin, we use {\sc redback} to fit the general magnetar-driven kilonova model from \citet{sarin22} to the data of \grb{} (Fig.~\ref{fig:redbackfit_magdrivenKN_split} and Fig.~\ref{fig:redbackfit_magdrivenKN}). Although the $r$-band data largely falls within the $90$\,per cent credible interval of the magnetar-driven kilonova model, \grb{} is brighter in $K$ and $z$ than expected for such a transient. Additionally, the maximum posterior light curve is a poor fit in both $K$ and $r$.

\subsection{{\em Fermi}/LAT Emission}

Our analysis of the {\em Fermi} data over the first $4$\,s after the burst gives a hard LAT spectral index of $-1.64 \pm 0.139$ and a significantly softer high-energy Band index derived from GBM data $\beta \approx -2.86\pm0.16$. This indicates the presence of an additional, harder, spectral component at GeV energies. Our broadband afterglow modelling (Sec.~\ref{sec:redback}; based on X-ray, optical, and IR data), including inverse Compton scattering, fails to reproduce the observed LAT flux (see Fig. \ref{fig:redbackfit_csm} and Fig. \ref{fig:redbackfit_shockcooling}). A simple extrapolation of the fitted afterglow model to GeV energies also underpredicts the LAT flux. The origin of the LAT emission therefore remains unclear. A detailed investigation of the hard GeV component lies beyond the scope of this work.

\subsection{SED Fitting Caveats}

We note that if an SN is present in the {\em HST} epochs, this will impact the accuracy of our SED fits. The {\em HST} and Gemini epochs that were used for the fitting would include the transient, therefore resulting in an overestimation of the brightness of the underlying galaxy. For consistency, we ran {\sc prospector} again with a fixed $z=2.34$ on adapted photometry, where we assumed only half of the flux of the late-time {\em HST} and Gemini data is attributed to the galaxy. The resulting galaxy parameters are consistent with the parameters that we obtained before, so the effect of a possible SN on our SED fits is small.

\section{Conclusions}\label{sec:conclusion}

We presented a detailed, multi-wavelength analysis of \grb{}. From our low-S/N afterglow spectrum, we determined a tentative redshift of z=2.34. Line stacking the spectrum and SED fitting the photometry of the underlying extended source corroborate this redshift, which we therefore deem to be the most likely redshift of the burst.

We consider the most likely progenitor of \grb{} to be a collapsar. Light curve modelling shows potential late-time SN emission. Moreover, {\em HST} imaging taken $58$~rest-frame days post-burst shows potential SN emission on top of the host galaxy. Additionally, SED fitting yields host galaxy properties typical for core-collapse GRB progenitors and the spectral lag of \grb{}, its isotropic-equivalent energy, its placement on the Amati relation, and its location on top of a galaxy all point to a core-collapse progenitor.

\grb{} can be classified as an EE-GRB, as its short initial pulse is followed by weaker, softer prompt emission. A collapsar origin is highly unusual for EE-GRBs, which are typically linked to compact object mergers. Nevertheless, proposed mechanisms for the extended emission such as magnetar spin-down or fallback accretion still apply if a (short-lived) magnetar remnant is formed after the SN.\\ 

The light curve of \grb{} contains rapid, highly luminous optical/IR emission that peaks around $7$\,d reaching $M_r=-22$\,mag. It resembles LFBOT emission, despite the LFBOT population generally not exhibiting obvious SN signatures at any epoch. Additionally, it closely resembles the bright emission peak observed in FXT EP241021a and has some similarities to the bright peak in FXT EP240414a. These similarities reinforce the notion that the same emission mechanisms could be at play for some LFBOTs, FXTs, and GRBs. Markedly, \grb{} has high-energy gamma-ray detections in {\em Fermi}/LAT, which indicates a high bulk Lorentz factor. The link between low-luminosity GRBs, FXTs, and LFBOTs has been suggested before (e.g. \citealt{vandalen25}), but \grb{} shows that high-luminosity GRBs could also be linked to these sources.

\citet{becerra23} disregard a collapsar origin at $z=2.34$ on the basis of not being able to explain the rapid, highly luminous optical/IR emission of \grb{}. However, our light curve modelling, using Bayesian inference package {\sc redback}, shows that the excess emission could potentially originate from an energetic refreshed shock in the GRB afterglow. As refreshed shocks can also be at play in merger GRBs (e.g. \citealt{lamb20}), the connection between \grb{}, FXTs, and LFBOTs does not necessarily suggest similar progenitors.

\section*{Acknowledgements}

This publication is part of the Dutch Black Hole Consortium with project number NWA.1292.19.202 of the research programme NWA which is (partly) financed by the Dutch Research Council (NWO).

J. C. Rastinejad was supported by NASA through the NASA Hubble Fellowship grant \#HST-HF2-51587.001-A awarded by the Space Telescope Science Institute, which is operated by the Association of Universities for Research in Astronomy, Inc., for NASA, under contract NAS5-26555.
A. Rossi acknowledges support from INAF project \textit{Supporto Arizona \& Italia}.
Dimple acknowledges support from STFC grant No. ST/Y002253/1.
DBM is funded by the European Union (ERC, HEAVYMETAL, 101071865). Views and opinions expressed are, however, those of the authors only and do not necessarily reflect those of the European Union or the European Research Council. Neither the European Union nor the granting authority can be held responsible for them. The Cosmic Dawn Center (DAWN) is funded by the Danish National Research Foundation under grant DNRF140.

Based on observations obtained at the international Gemini Observatory (programme ID GN-2021B-Q-109), a programme of NOIRLab, which is managed by the Association of Universities for Research in Astronomy (AURA) under a cooperative agreement with the National Science Foundation on behalf of the Gemini Observatory partnership: the National Science Foundation (United States), National Research Council (Canada), Agencia Nacional de Investigaci\'{o}n y Desarrollo (Chile), Ministerio de Ciencia, Tecnolog\'{i}a e Innovaci\'{o}n (Argentina), Minist\'{e}rio da Ci\^{e}ncia, Tecnologia, Inova\c{c}\~{o}es e Comunica\c{c}\~{o}es (Brazil), and Korea Astronomy and Space Science Institute (Republic of Korea). Processed using the Gemini {\sc iraf} package and {\sc dragons}.

This work is partly based on data obtained with the instrument OSIRIS, built by a Consortium led by the Instituto de Astrof\'isica de Canarias in collaboration with the Instituto de Astronom\'ia of the Universidad Aut\'onoma de M\'exico. OSIRIS was funded by GRANTECAN and the National Plan of Astronomy and Astrophysics of the Spanish Government. The GTC data were obtained under programme GTCMULTIPLE2C-21A.

This work is partly based on observations made with the Nordic Optical Telescope, owned in collaboration by the University of Turku and Aarhus University, and operated jointly by Aarhus University, the University of Turku and the University of Oslo, representing Denmark, Finland and Norway, the University of Iceland and Stockholm University at the Observatorio del Roque de los Muchachos, La Palma, Spain, of the Instituto de Astrofisica de Canarias. The NOT data were obtained under programme ID 63-503.

This publication is partly based on observations made with the Javalambre Observatory (OAJ), within programme 2100193 (PI: Ag\"u\'i Fern\'andez, J. F.).

Partly based on observations made with the Italian Telescopio Nazionale Galileo (TNG) operated by the Fundación Galileo Galilei (FGG) of the Istituto Nazionale di Astrofisica (INAF) at the Observatorio del Roque de los Muchachos (La Palma, Canary Islands, Spain). The TNG data were obtained under programme A43TAC\_2.

Partly based on observations made with the William Herschel Telescope operated on the island of La Palma by the Isaac Newton Group of Telescopes in the Spanish Observatorio del Roque de los Muchachos of the Instituto de Astrof\'isica de Canarias (proposal QHY).

Partly based on observations collected at Centro Astron\'omico Hispano en Andaluc\'ia (CAHA) at Calar Alto, proposal 21B-2.2-018, operated jointly by Junta de Andaluc\'ia and Consejo Superior de Investigaciones Cient\'ificas (IAA-CSIC).

This research is based on observations made with the NASA/ESA {\em Hubble Space Telescope} obtained from the Space Telescope Science Institute, which is operated by the Association of Universities for Research in Astronomy, Inc., under NASA contract NAS 5–26555. These observations are associated with programme 16275.

This work is partially based on observations obtained at the LBT under programme ID IT-2021B-018. The LBT is an international collaboration of the University of Arizona, Italy (INAF: Istituto Nazionale di Astrofisica), Germany (LBTB: LBT Beteiligungsgesellschaft), The Ohio State University, representing also the University of Minnesota, the University of Virginia, and the University of Notre Dame.

Based on observations obtained with MegaPrime/MegaCam, a joint project of CFHT and CEA/DAPNIA, at the Canada-France-Hawaii Telescope (CFHT) which is operated by the National Research Council (NRC) of Canada, the Institut National des Science de l'Univers of the Centre National de la Recherche Scientifique (CNRS) of France, and the University of Hawaii. The observations at the Canada-France-Hawaii Telescope were performed with care and respect from the summit of Maunakea which is a significant cultural and historic site. 

This paper is based (in part) on data from the Hyper Suprime-Cam Legacy Archive (HSCLA), which is operated by the Subaru Telescope. The original data in HSCLA were collected at the Subaru Telescope and retrieved from the HSC data archive system, which is operated by the Subaru Telescope and Astronomy Data Center at National Astronomical Observatory of Japan. The Subaru Telescope is honoured and grateful for the opportunity of observing the Universe from Maunakea, which has the cultural, historical, and natural significance in Hawaii.

This paper makes use of software developed for the Vera C. Rubin Observatory. We thank the observatory for making their code available as free software at  \url{http://dm.lsst.org}.

The Pan-STARRS1 Surveys (PS1) and the PS1 public science archive have been made possible through contributions by the Institute for Astronomy, the University of Hawaii, the Pan-STARRS Project Office, the Max-Planck Society and its participating institutes, the Max Planck Institute for Astronomy, Heidelberg and the Max Planck Institute for Extraterrestrial Physics, Garching, The Johns Hopkins University, Durham University, the University of Edinburgh, the Queen's University Belfast, the Harvard-Smithsonian Center for Astrophysics, the Las Cumbres Observatory Global Telescope Network Incorporated, the National Central University of Taiwan, the Space Telescope Science Institute, the National Aeronautics and Space Administration under Grant No. NNX08AR22G issued through the Planetary Science Division of the NASA Science Mission Directorate, the National Science Foundation Grant No. AST-1238877, the University of Maryland, Eotvos Lorand University (ELTE), the Los Alamos National Laboratory, and the Gordon and Betty Moore Foundation.

This publication makes use of data products from the Two Micron All Sky Survey, which is a joint project of the University of Massachusetts and the Infrared Processing and Analysis Center/California Institute of Technology, funded by the National Aeronautics and Space Administration and the National Science Foundation.

Funding for the Sloan Digital Sky Survey V has been provided by the Alfred P. Sloan Foundation, the Heising-Simons Foundation, the National Science Foundation, and the Participating Institutions. SDSS acknowledges support and resources from the Center for High-Performance Computing at the University of Utah. SDSS telescopes are located at Apache Point Observatory, funded by the Astrophysical Research Consortium and operated by New Mexico State University, and at Las Campanas Observatory, operated by the Carnegie Institution for Science. The SDSS web site is \url{www.sdss.org}.

SDSS is managed by the Astrophysical Research Consortium for the Participating Institutions of the SDSS Collaboration, including the Carnegie Institution for Science, Chilean National Time Allocation Committee (CNTAC) ratified researchers, Caltech, the Gotham Participation Group, Harvard University, Heidelberg University, The Flatiron Institute, The Johns Hopkins University, L'Ecole polytechnique f\'{e}d\'{e}rale de Lausanne (EPFL), Leibniz-Institut f\"{u}r Astrophysik Potsdam (AIP), Max-Planck-Institut f\"{u}r Astronomie (MPIA Heidelberg), Max-Planck-Institut f\"{u}r Extraterrestrische Physik (MPE), Nanjing University, National Astronomical Observatories of China (NAOC), New Mexico State University, The Ohio State University, Pennsylvania State University, Smithsonian Astrophysical Observatory, Space Telescope Science Institute (STScI), the Stellar Astrophysics Participation Group, Universidad Nacional Aut\'{o}noma de M\'{e}xico, University of Arizona, University of Colorado Boulder, University of Illinois at Urbana-Champaign, University of Toronto, University of Utah, University of Virginia, Yale University, and Yunnan University. 

This work made use of {\sc astropy}:\footnote{http://www.astropy.org} a community-developed core Python package and an ecosystem of tools and resources for astronomy~\citep{astropy22, astropy24}. Additionally, this work made use of {\sc astroquery}~\citep{ginsburg19, ginsburg25}, {\sc matplotlib}~\citep{hunter07, matplotlib25}, {\sc numpy}~\citep{harris20}, and {\sc pandas}~\citep{mckinney10, pandas25}.

\section*{Data Availability}
The data underlying this article are available in the article.


\bibliographystyle{mnras}
\bibliography{grb210704a}


\appendix
\section{Corner plots}

\begin{figure*}
    \centering
    \includegraphics[width=0.65\textwidth]{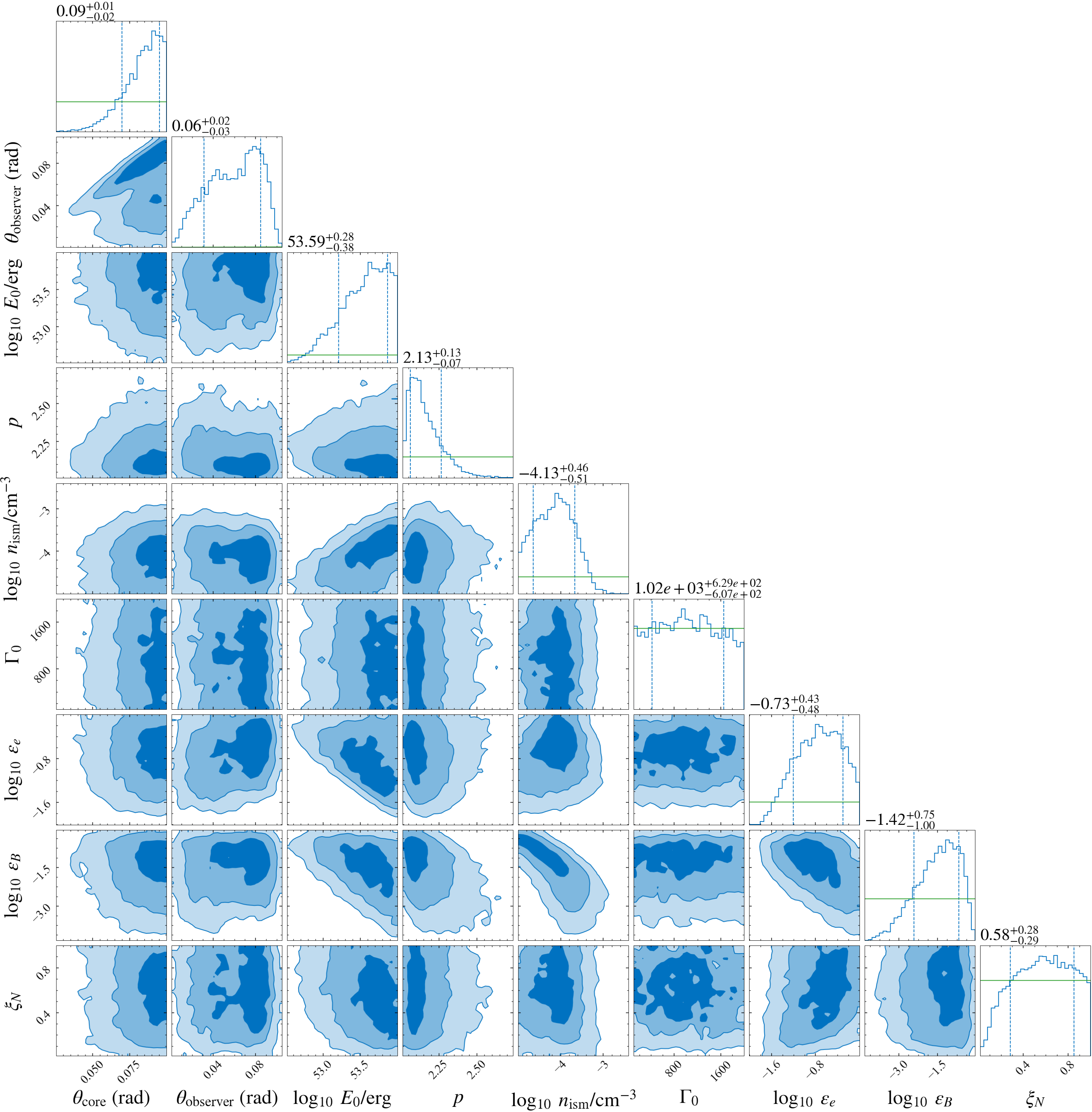}\\
    \includegraphics[width=0.43\textwidth]{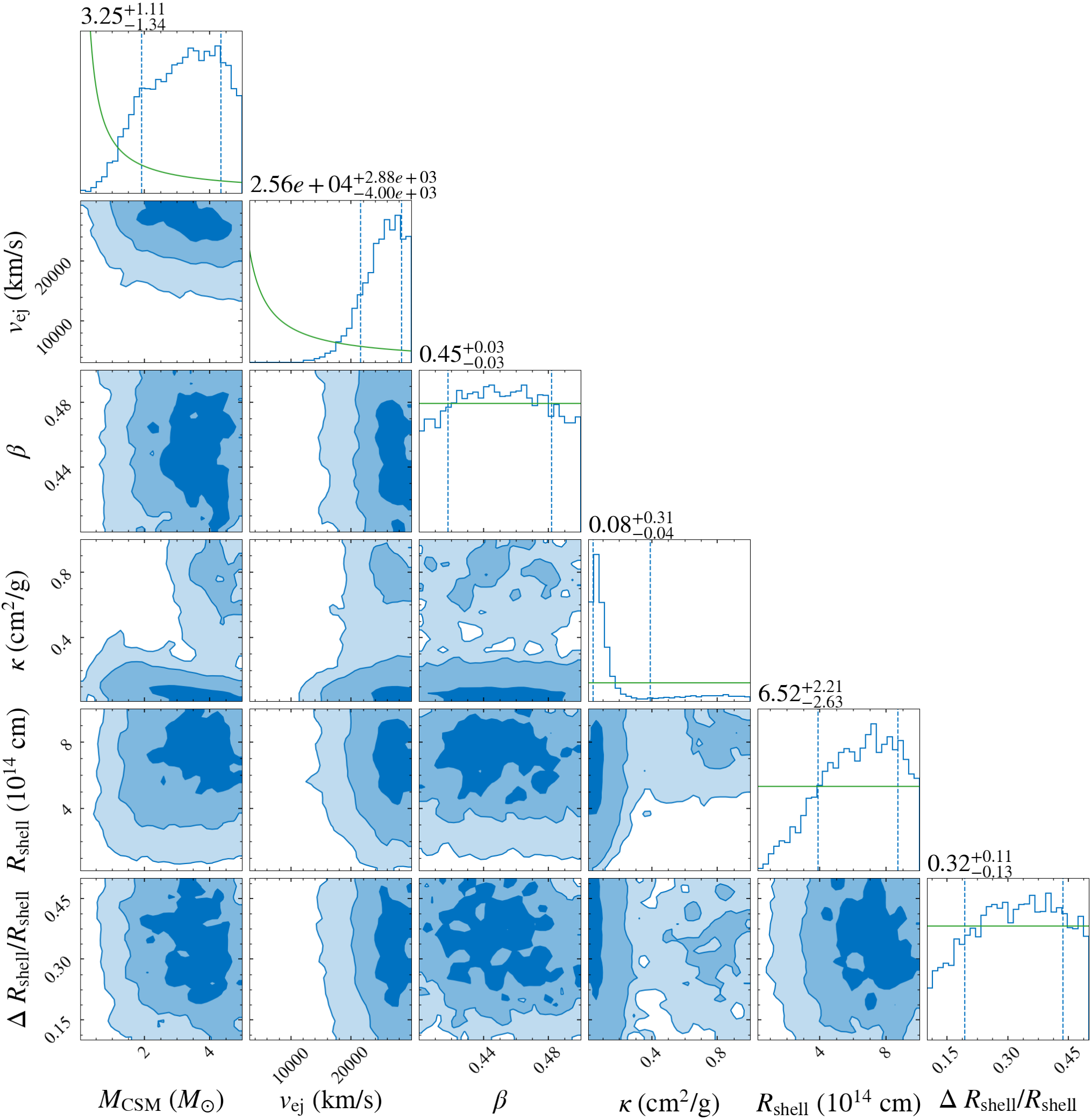}
    \includegraphics[width=0.43\textwidth]{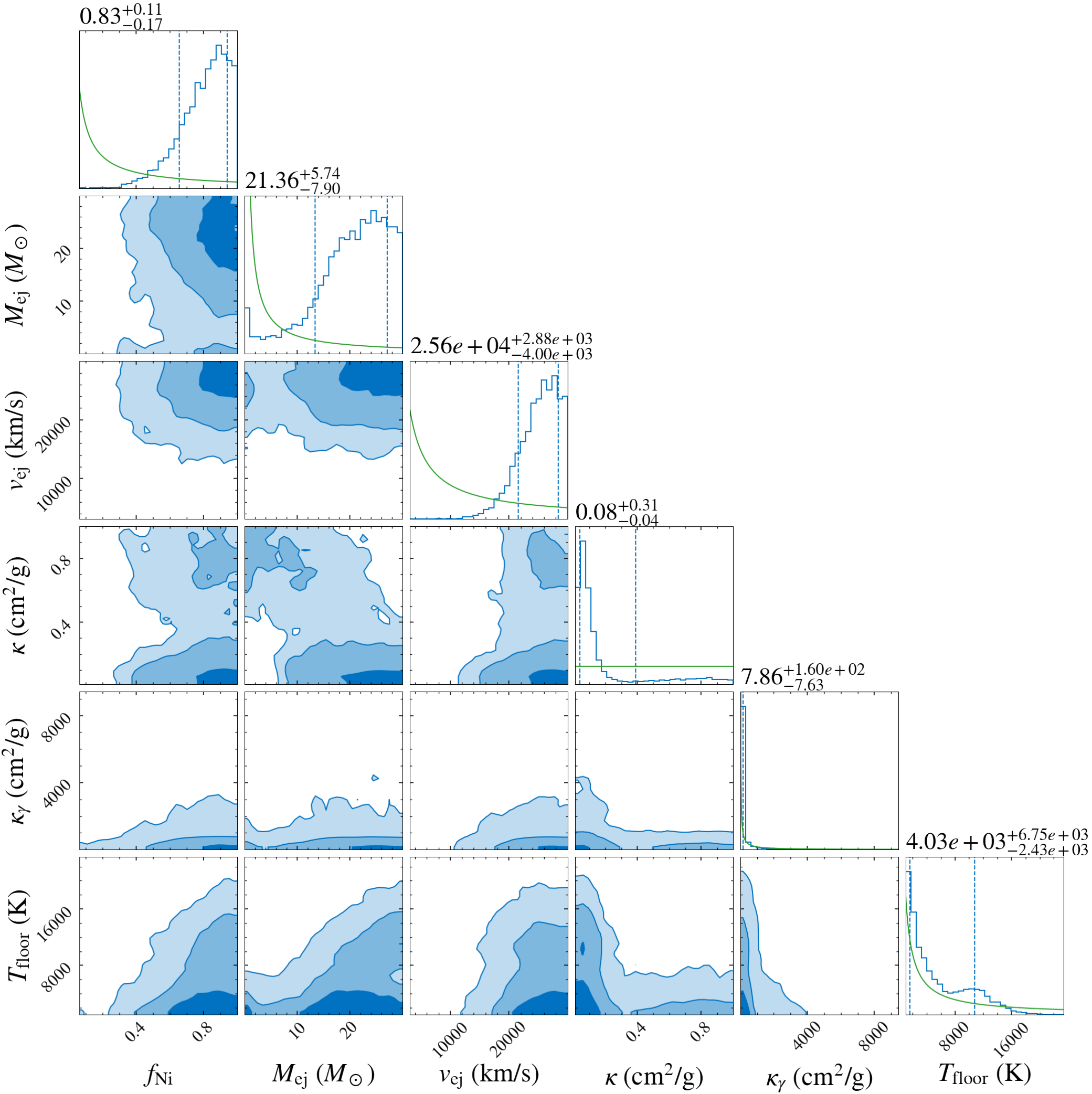}
    \caption{Posterior distribution from the joint fit using a top-hat afterglow model (upper plot; \protect\citealt{ryan20}), an SN-CSM interaction model (lower left plot; \protect\citealt{margalit22}), and an SN model (lower right plot; \protect\citealt{arnett82}). The graphs in the histogram panels correspond to the prior distributions. The parameter values above the histograms are the median, where the 16th and 84th percentiles represent the error bars (also indicated by the vertical dashed lines in the histograms). The posterior parameters for the top-hat model are the jet opening angle $\theta_{\text{core}}$, viewing angle $\theta_{\text{observer}}$, jet energy $E_0$, electron power law index $p$, CSM number density $n_{\text{ism}}$, initial Lorentz factor $\Gamma_0$, partition fractions in electrons ($\epsilon_e$) and in magnetic field ($\epsilon_B$), and the fraction of electrons that get accelerated $\xi_N$. For the CSM interaction model, the parameters are the CSM-shell mass $M_{\text{CSM}}$, minimum ejecta velocity $v_{\text{ej}}$, velocity ratio $\beta=v/c$, effective opacity $\kappa$, and the radius and width of the shell ($R_{\text{shell}}$ and $\Delta R_{\text{shell}}$). For the SN model, the parameters are the nickel mass fraction $f_{\text{Ni}}$, total ejecta mass $M_{\text{ej}}$, opacity to gamma rays $\kappa_{\gamma}$, temperature floor $T_{\text{floor}}$, and $v_{\text{ej}}$ and $\kappa$ that are also in the CSM interaction model.}
    \label{fig:corner_csm}
\end{figure*}

\begin{figure*}
    \centering
    \includegraphics[width=0.6\textwidth]{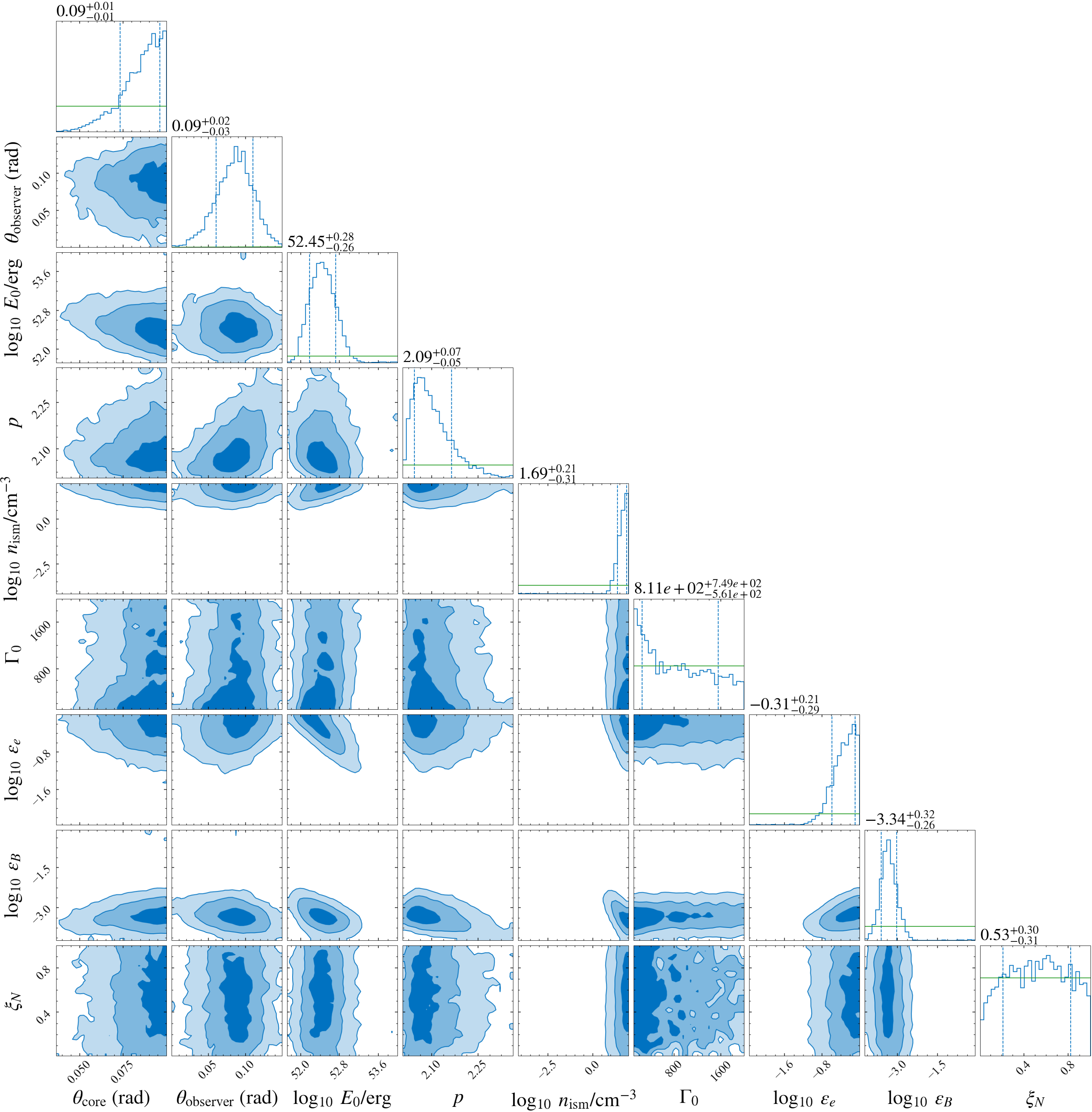}
    \includegraphics[width=0.22\textwidth]{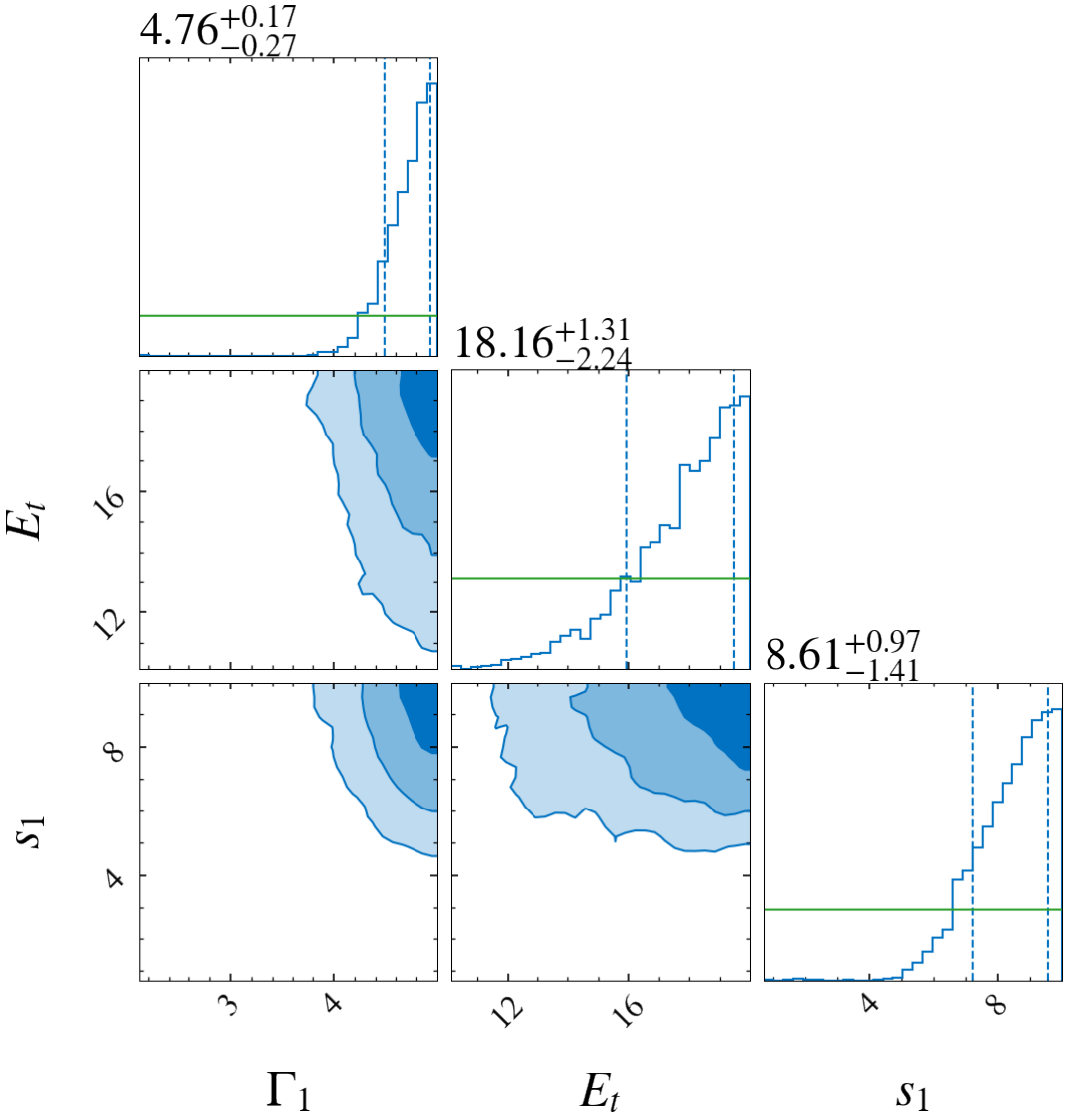}
    \includegraphics[width=0.43\textwidth]{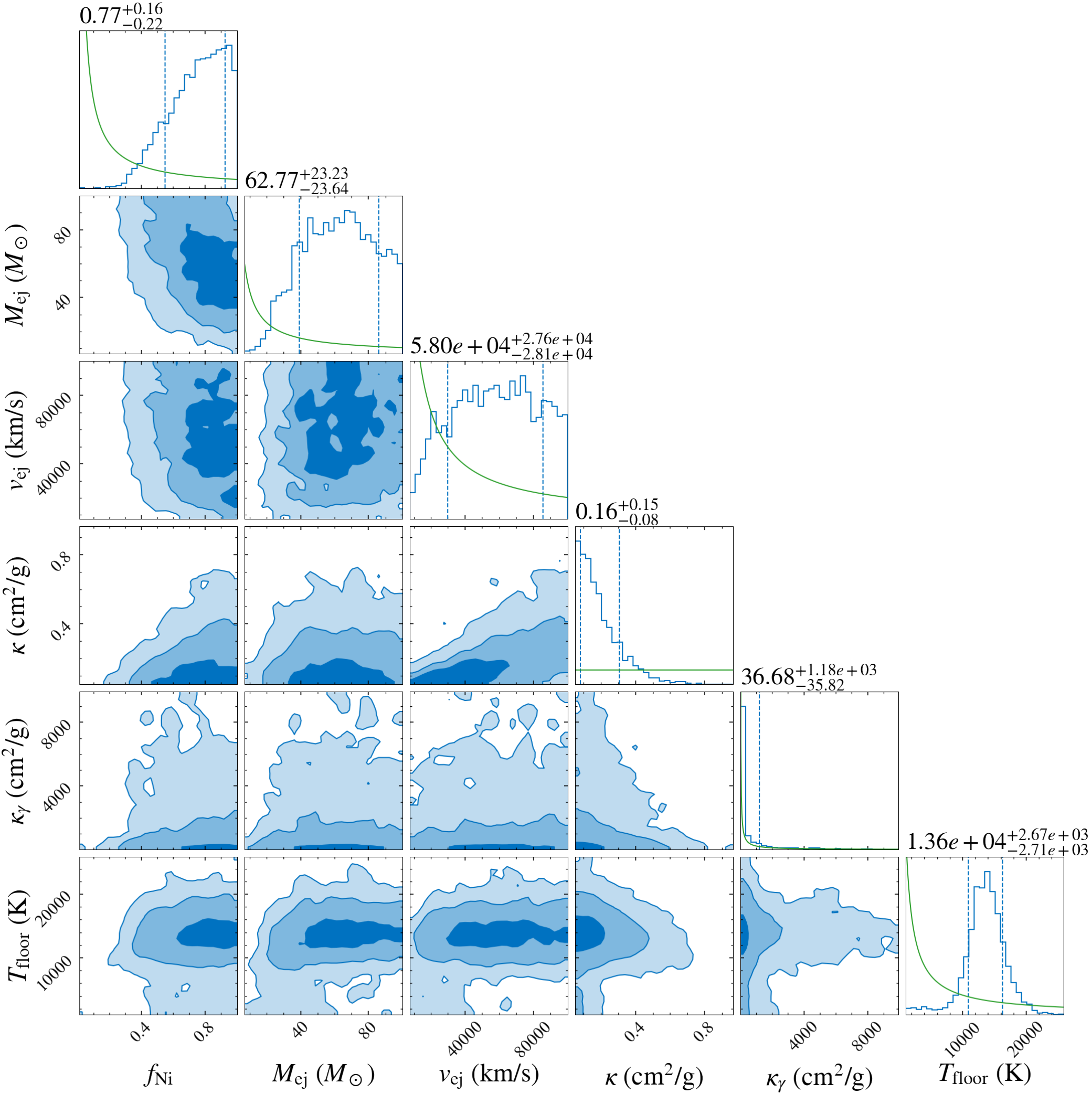}
    \caption{Posterior distribution from the joint fit using a refreshed top-hat afterglow model (upper left plot for the standard afterglow parameters and upper right plot for the refreshed shock parameters; \protect\citealt{lamb19, lamb20}) and an SN model (lower plot; \protect\citealt{arnett82}). The graphs in the histogram panels correspond to the prior distributions. The parameter values above the histograms are the median, where the 16th and 84th percentiles represent the error bars (also indicated by the vertical dashed lines in the histograms). The posterior parameters for the top-hat model are the jet opening angle $\theta_{\text{core}}$, viewing angle $\theta_{\text{observer}}$, jet energy $E_0$, electron power law index $p$, CSM number density $n_{\text{ism}}$, initial Lorentz factor $\Gamma_0$, partition fractions in electrons ($\epsilon_e$) and in magnetic field ($\epsilon_B$), and the fraction of electrons that get accelerated $\xi_N$. For the refreshed shock, the parameters are the Lorentz factor of the shell at the start of energy injection $\Gamma_1$, the factor by which the kinetic energy is larger $E_t$, and the index of energy injection $s_1$. For the SN model, the parameters are the nickel mass fraction $f_{\text{Ni}}$, total ejecta mass $M_{\text{ej}}$, minimum ejecta velocity $v_{\text{ej}}$, effective opacity $\kappa$, opacity to gamma rays $\kappa_{\gamma}$, and the temperature floor $T_{\text{floor}}$.}
    \label{fig:corner_refreshed}
\end{figure*}

\begin{figure*}
    \centering
    \includegraphics[width=0.65\textwidth]{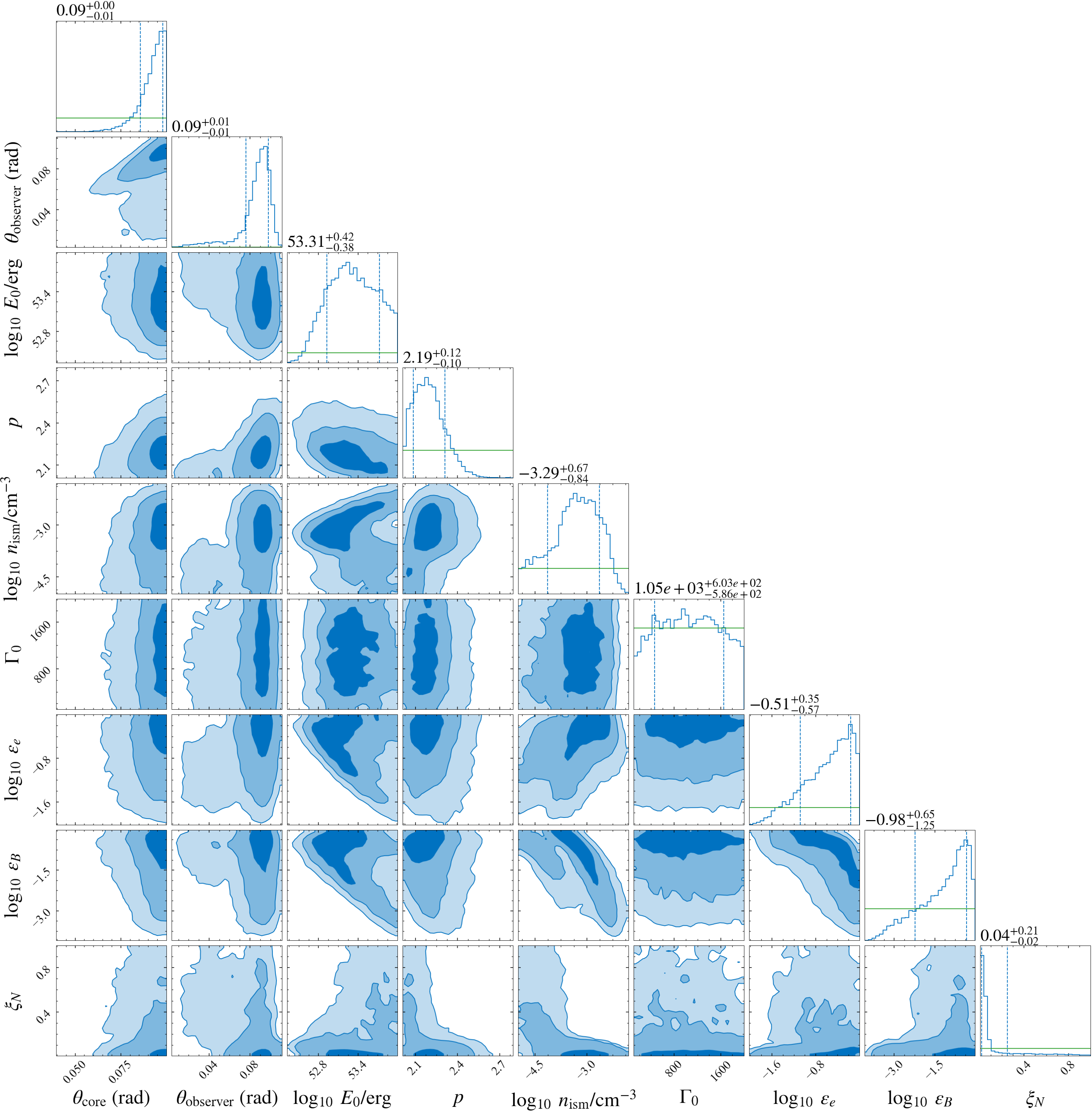}\\
    \includegraphics[width=0.36\textwidth]{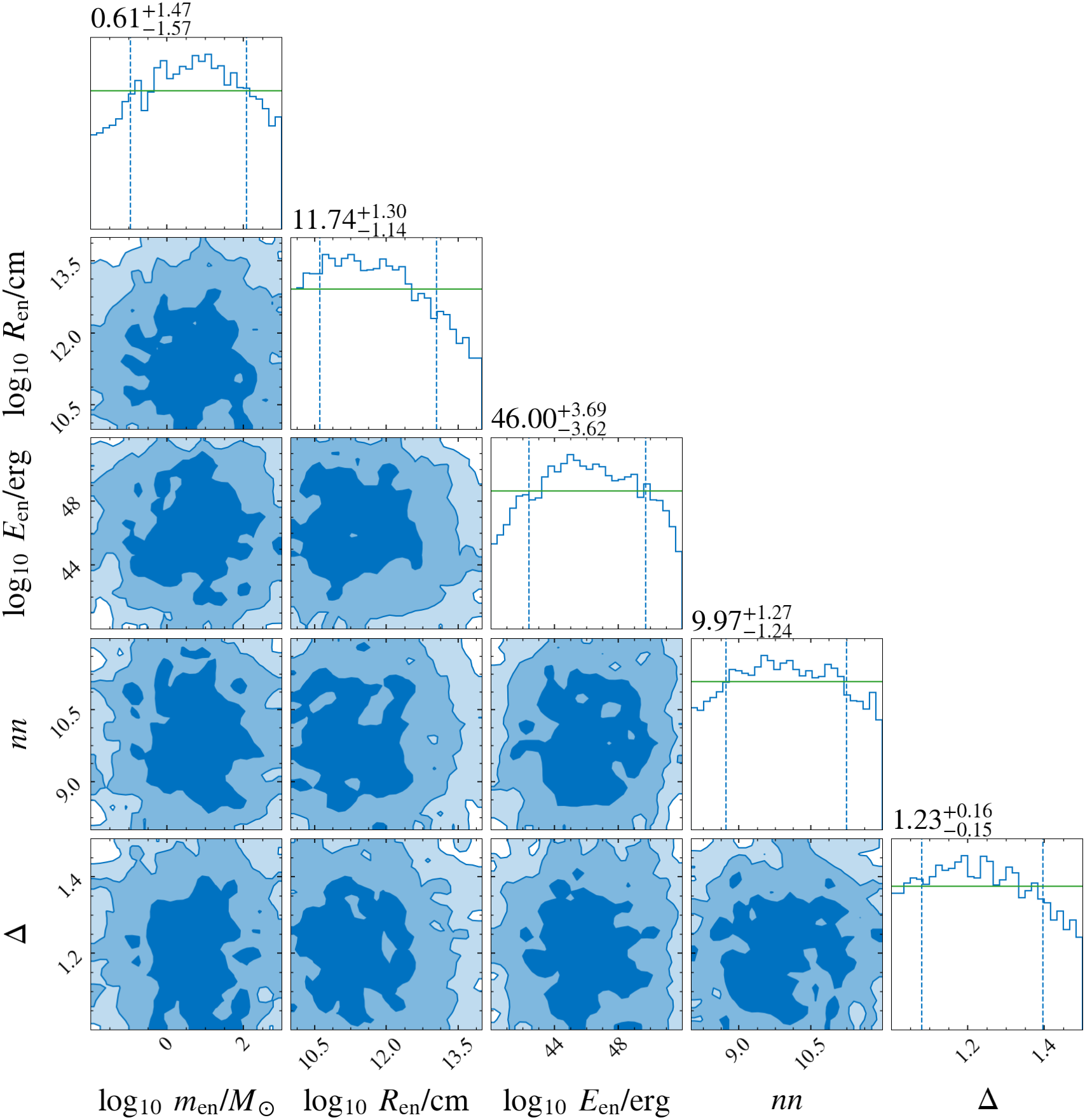}
    \includegraphics[width=0.43\textwidth]{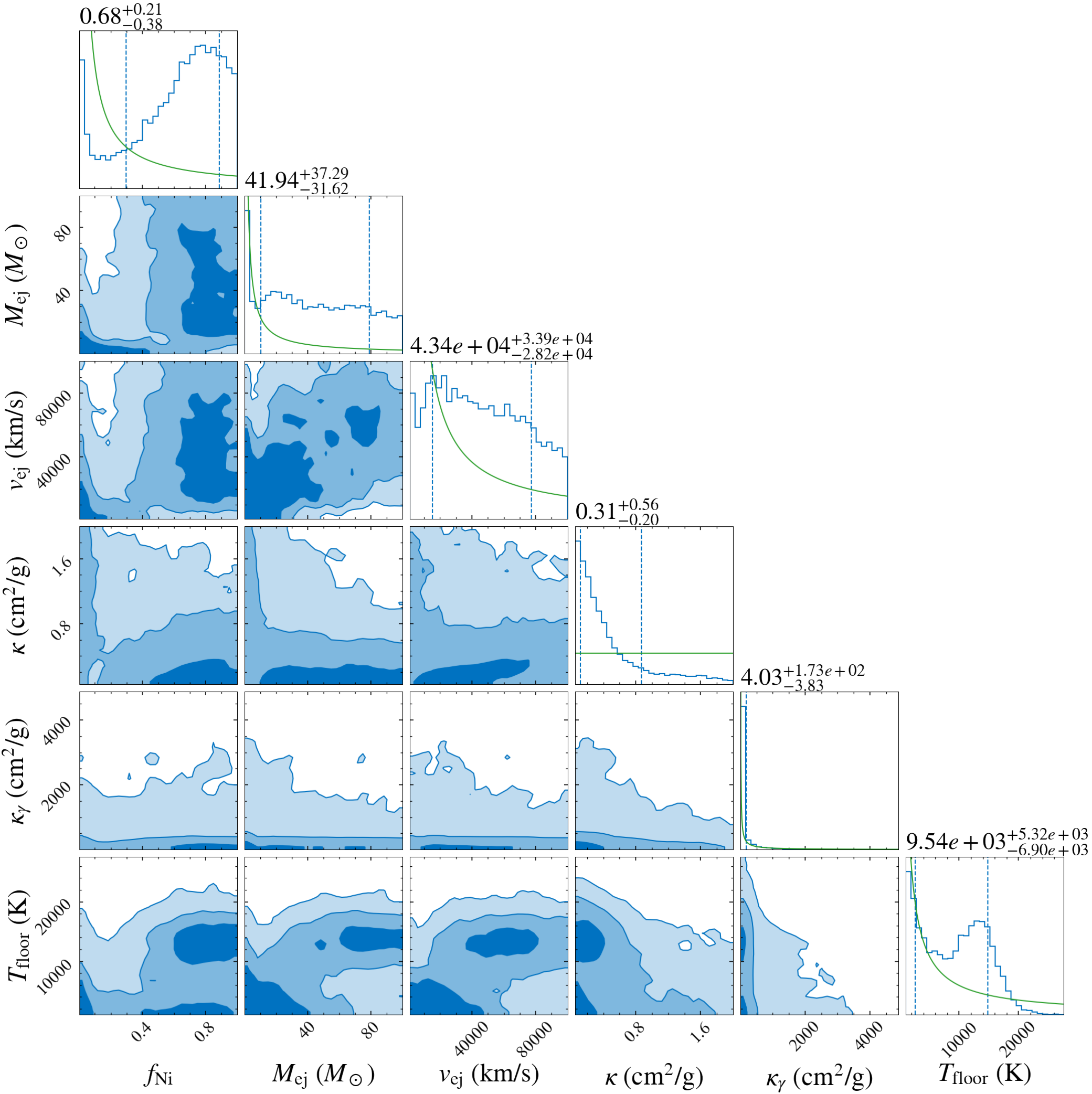}
    \caption{Posterior distribution from the joint fit using a top-hat afterglow model (upper plot; \protect\citealt{ryan20}), shock cooling model (lower left plot; \protect\citealt{piro21}) and an SN model (lower right plot; \protect\citealt{arnett82}). The graphs in the histogram panels correspond to the prior distributions. The parameter values above the histograms are the median, where the 16th and 84th percentiles represent the error bars (also indicated by the vertical dashed lines in the histograms). The posterior parameters for the top-hat model are the jet opening angle $\theta_{\text{core}}$, viewing angle $\theta_{\text{observer}}$, jet energy $E_0$, electron power law index $p$, CSM number density $n_{\text{ism}}$, initial Lorentz factor $\Gamma_0$, partition fractions in electrons ($\epsilon_e$) and in magnetic field ($\epsilon_B$), and the fraction of electrons that get accelerated $\xi_N$. For the shock cooling model, the parameters are the mass $m_{\text{en}}$, energy $E_{\text{en}}$, and radius $R_{\text{en}}$ of the extended material and the power-law density slopes of the outer and inner extended material components ($nn$ and $\Delta$, respectively). For the SN model, the parameters are the nickel mass fraction $f_{\text{Ni}}$, total ejecta mass $M_{\text{ej}}$, minimum ejecta velocity $v_{\text{ej}}$, effective opacity $\kappa$, opacity to gamma rays $\kappa_{\gamma}$, and the temperature floor $T_{\text{floor}}$.}
    \label{fig:corner_shockcooling}
\end{figure*}

\onecolumn
\section*{Affiliations}
$^{1}$Department of Astrophysics/IMAPP, Radboud University Ni\nobreak{}jmegen, P.O.~Box 9010, Nijmegen, 6500~GL, The Netherlands\\
$^{2}$Department of Physics, University of Warwick, Coventry, CV4~7AL, UK\\
$^{3}$INAF -- Osservatorio Astronomico di Brera, Via E. Bianchi 46, Merate, I-23807, Italy\\
$^{4}$Center for Interdisciplinary Exploration and Research in Astrophysics (CIERA) and Department of Physics and Astronomy, Northwestern University, Evanston, IL~60208, USA\\
$^{5}$Department of Astronomy, University of Maryland, College Park, MD~20742, USA\\
$^{6}$Cosmic Dawn Center (DAWN), Denmark\\
$^{7}$Niels Bohr Institute, University of Copenhagen, Jagtvej~155, 2200~Copenhagen~N, Denmark\\
$^{8}$Kavli Institute for Cosmology Cambridge, Madingley Road, Cambridge CB3~0HA, United Kingdom\\
$^{9}$Institute of Astronomy, University of Cambridge, Madingley Road, Cambridge CB3~0HA, United Kingdom\\
$^{10}$Astrophysics Research Institute, Liverpool John Moores University, IC2~Liverpool Science Park, 146~Brownlow Hill, Liverpool, L3~5RF, UK\\
$^{11}$School of Physics and Centre for Space Research, O’Brien Centre for Science North, University College Dublin, Belfield, Dublin~4, Ireland\\
$^{12}$Center for Astrophysics\:$|$\:Harvard \& Smithsonian, 60~Garden St., Cambridge, MA~02138, USA\\
$^{13}$School of Physics and Astronomy, University of Leicester, University Road, Leicester, LE1~7RH, United Kingdom\\
$^{14}$Hessian Research Cluster ELEMENTS, Giersch Science Center, Max-von-Laue-Strasse~12, Goethe University Frankfurt, Campus Riedberg, 60438~Frankfurt am Main, Germany\\
$^{15}$Instituto de Astrof\'isica de Andaluc\'ia (IAA-CSIC), Glorieta de la Astronom\'ia s/n, E-18008 Granada, Spain\\
$^{16}$Centro Astron\'omico Hispano en Andaluc\'ia, Observatorio de Calar Alto, Sierra de los Filabres, G\'ergal, Almer\'ia, 04550, Spain\\
$^{17}$INAF -- Osservatorio di Astrofisica e Scienza dello Spazio, Via Piero Gobetti~93/3, I-40024, Bologna, Italy\\
$^{18}$Space Science Data Center (SSDC) -- Agenzia Spaziale Italiana (ASI), 00133~Roma, Italy\\
$^{19}$Aix Marseille University, CNRS, CNES, LAM, Marseille, France\\
$^{20}$School of Physics and Astronomy, University of Birmingham, Edgbaston,
Birmingham B15 2TT, UK\\
$^{21}$Institute for Gravitational Wave Astronomy, University of Birmingham,
Birmingham B15 2TT, UK\\
$^{22}$INAF -- Osservatorio Astronomico di Capodimonte, Salita Moiariello~16, I-80131, Naples, Italy\\
$^{23}$DARK, Niels Bohr Institute, University of Copenhagen, Jagtvej~128, 2200~Copenhagen, Denmark\\
$^{24}$INAF -- Osservatorio Astronomico di Roma, Via di Frascati~33, I-00040, Monte Porzio Catone (RM), Italy\\
$^{25}$European Space Agency (ESA), European Space Astronomy Centre (ESAC), Camino Bajo del Castillo s/n, 28692 Villanueva de la Ca\~nada, Madrid, Spain\\

\bsp	
\label{lastpage}
\end{document}